\documentclass{vgtc}                          % final (conference style)
\ifpdf%                                % if we use pdflatex
  \pdfoutput=1\relax                   % create PDFs from pdfLaTeX
  \usepackage{graphicx}                % allow us to embed graphics files
  \DeclareGraphicsExtensions{.pdf,.png,.jpg,.jpeg} % for pdflatex we expect .pdf, .png, or .jpg files
\else%                                 % else we use pure latex
  \usepackage{graphicx}                % allow us to embed graphics files
  \DeclareGraphicsExtensions{.eps}     % for pure latex we expect eps files
\fi%

\graphicspath{{figures/}{pictures/}{images/}{./}} % where to search for the images

\usepackage{microtype}                 % use micro-typography (slightly more compact, better to read)
\PassOptionsToPackage{warn}{textcomp}  % to address font issues with \textrightarrow
\usepackage{textcomp}                  % use better special symbols
\usepackage{mathptmx}                  % use matching math font
\usepackage{times}                     % we use Times as the main font
\usepackage{cite}                      % needed to automatically sort the references
\usepackage{tabu}                      % only used for the table example
\usepackage{booktabs}                  % only used for the table example
\onlineid{0}

\vgtccategory{Research}

\vgtcinsertpkg

\preprinttext{To appear in an IEEE VGTC sponsored conference.}

\title{Visual analytics for networked-guarantee loans risk management}

\author{Zhibin Niu\thanks{e-mail: zniu@tju.edu.cn}\\ %
        \scriptsize Tianjin University %
\and Dawei Cheng\thanks{e-mail: dawei.cheng@sjtu.edu.cn}\\ %
{\scriptsize \centering Shanghai Jiao Tong University}% \parbox{1.4in}
\and Liqing Zhang\thanks{e-mail: lq-zhang@sjtu.edu.cn }\\ %\parbox{1.4in}
{\scriptsize \centering Shanghai Jiao Tong University}
\and Jiawan Zhang\thanks{e-mail: jwzhang@tju.edu.cn}\\ %
     \scriptsize Tianjin University}

\abstract{Groups of enterprises can guarantee each other and form complex networks in order to try to obtain loans from banks. %On the one hand, secured loans such as these can enhance solvency and promote rapid growth in a period of economic upturn; on the other hand, a debt crisis in the network may spread like a domino effect within the network during a period of economic downturn.
Monitoring the financial status of a network, and preventing or reducing systematic risk in case of a crisis, is an area of great concern for the regulatory commission and for the banks. We set the ultimate goal of developing a visual analytic approach and tool for risk dissolving and decision-making. We have consolidated four main analysis tasks conducted by financial experts: i) Multi-faceted Default Risk Visualization, whereby a hybrid representation is devised to predict the default risk and an interface developed to visualize key indicators; ii) Risk Guarantee Patterns Discovery. We follow the Shneiderman mantra guidance for designing interactive visualization applications, whereby an interactive risk guarantee community detection and a motif detection based risk guarantee pattern discovery approach are described; iii) Network Evolution and Retrospective, whereby animation is used to help users to understand the guarantee dynamic; iv) Risk Communication Analysis. The temporal diffusion path analysis can be useful for the government and banks to monitor the spread of the default status. It also provides insight for taking precautionary measures to prevent and dissolve systematic financial risk. We implement the system with case studies using real-world bank loan data. Two financial experts are consulted to endorse the developed tool. To the best of our knowledge, this is the first visual analytics tool developed to explore networked-guarantee loan risks in a systematic manner.%
} % end of abstract

\CCScatlist{
  \CCScat{H.5.2}{User Interfaces}{User Interfaces}{Graphical user interfaces (GUI)}{};
  \CCScat{H.5.m}{Information Interfaces and Presentation}{Miscellaneous}{}{}
}

\begin{document}


\firstsection{Introduction}

\maketitle

\begin{figure}[tb!]\vspace{-10pt}
  \centering
  \includegraphics[width=1\linewidth]{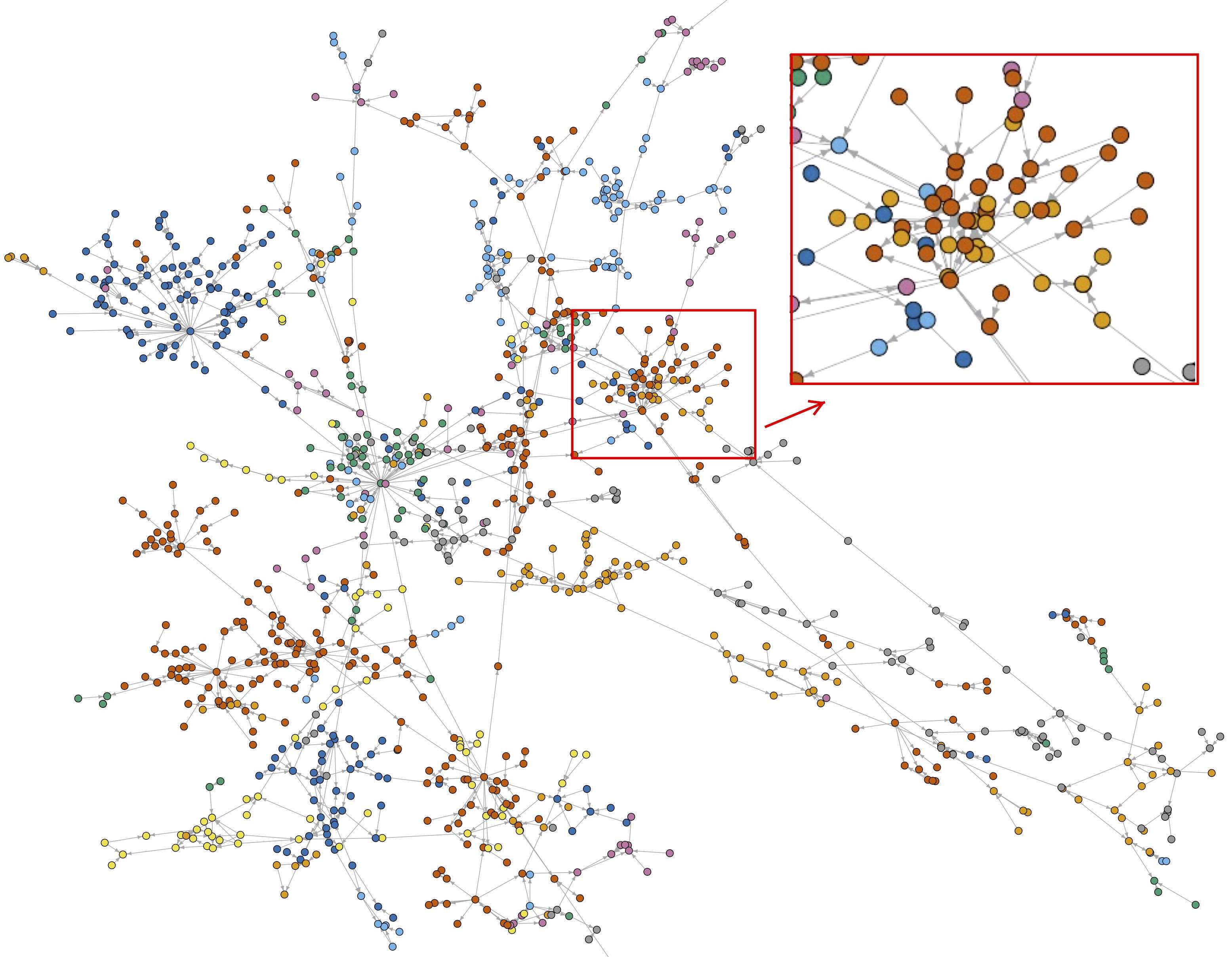}\vspace{-10pt}
  \caption{A real-world loan guarantee network formed from bank records, with each node representing an enterprise. }\label{realnetwork}\vspace{-19pt}
\end{figure}

Networked-guarantee loans (also known as guarantee circles) are an economic phenomenon unique to Asia countries, especially China, and they are attracting increasing attention from the banks and the government. In order to obtain loans from banks, groups of small and medium enterprises back each other to enhance their financial security. When more and more enterprises are involved, they form complex directed-network structures~\cite{meng2017netrating}. Figure~\ref{realnetwork} shows a guarantee network consisting of more than 600 enterprises. The existing mechanism in the financial industry for loan decision-making falls behind the demand for loans from businesses. Most of the criteria are designed for \emph{independent }\emph{major} players, while, in practice, the small and medium enterprises may provide inaccurate or manipulated data or induce intertwined risk factors~\cite{jian2012determinants}. Thousands of guarantee networks of different complexities have coexisted for a long period and have evolved over time. This requires an adaptive strategy in order to prevent, identify, and dismantle systematic crises.

Highlighted by the complex background of the growth period, the structural adjustment of the pain period, and the early stage of the stimulus period, structural and deep-level contradictions have emerged in the economic development system. Many kinds of risk factors have emerged throughout the guarantee network that might accelerate the transmission and amplification of risk, and the guarantee network may be alienated from the ``mutual aid group'' as a ``breach of contract''. An appropriate guarantee union may reduce the default risk, but significant contagious damage throughout the networked enterprises may still occur in practice~\cite{mcmahon2014loan}. The guaranteed loan is a debt obligation promise; if one corporation gets trapped in risks, it may spread the contagion to other corporations in the network. When defaults diffuse across the network, a systemic financial crisis may occur. The contagion to risk loan guarantee, especially malicious guarantee, is still relatively limited. Monitoring the financial status is so difficult that it is usually only after a capital chain rupture that the regulators can study a case in depth. With the economic slowdown, the need for credit risk management is more urgent than ever before.

We propose a visual analytics approach for networked-guarantee loan risk management. The main contributions are:

\begin{enumerate}\vspace{-5pt}
  \item   We identify and provide practical solution to the problem of credit risk management for networked-guarantee loans, which is driven by finance industry demands, and we believe this is an important research problem to the data mining and visual analytics community.\vspace{-5pt}
  \item We implement intuitive visual analytic tools for i) Multi-faceted Default Risk Visualization; ii)  Risk Guarantee Patterns Discovery; iii) Network Evolution and Retrospective; and iv) Risk Communication Analysis. We perform empirical studies and verified the efficacy. \vspace{-5pt}
  \item We conduct interviews with two domain experts and have our approach endorsed. We highlight three risk patterns that are difficult to discern without using a visual analytic approach.\vspace{-5pt}
\end{enumerate}

The rest of the paper is organized as following: Section 2 describes works involving different aspects related to our problem; Section 3 details the four visual analytic tasks and our approaches; Section 4 describes the data and the case study; and we report the user study results in Section 5. Conclusions and future works are described in Section 6.

\section{Related Work}
We introduce several relevant works on network analytics in the financial domain and works on financial security visualization.

\textbf{Credit risk evaluation} Since the seminal ``Partial Credit'' model~\cite{masters1982rasch}, numerous data-driven approaches have been introduced for credit scoring~\cite{baesens2003benchmarking}. Jan Vanthienen and others interpreted and visualized the learned knowledge embedded in neural networks based credit scoring approach~\cite{baesens2003using}. Andrew W. Lo and others propose consumer credit risk prediction models based on consumer behavior (debt-to-income ratio and consumer banking transactions), linear regression model, and time-windowed data set. They claim a 85\% default prediction accuracy and can save cost between 6\% and 25\%~\cite{khandani2010consumer}. In this paper, we adopt a similar idea and propose a hybrid representation to predict the enterprise default rate.

\textbf{Financial network analytics} The relationship between network structure and financial system risk has been studied carefully and several insights have been drawn: Network structure has little impact on system welfare, but plays an important role in determining systematic risk and welfare in the short term debt~\cite{allen2010financial}. After the 2008 global financial crisis, network theory attracted more attention: The crisis brought about by Lehman Brothers spread to connected corporations in a similar infectious way as the epidemic of Severe Acute Respiratory Syndrome (SARS) in 2002 -- both were small damages that hit a networked system and caused serious events~\cite{bougheas2015complex}. The journal of Nature Physics published a special edition on how to understand some fundamental economic issues using network theory. For example, the dynamic network produced by bank overnight fund loans may act as an alert of a crisis~\cite{catanzaro2013network}. Contrary to the conventional stereotype that large institutions are ``too big to fail'', the truth is that the position of an institution in a network is equally, and sometimes more, important than its size~\cite{battiston2012debtrank}. The more central the vertex is to the graph, the more influential it is to the whole economic network when default occurs~\cite{catanzaro2013network}. %Moreover, research that aims to understand individual behavior and interactions in the social network has also attracted extensive attention~\cite{zafarani2015evaluation,qiu2016lifecycle}.
Although considerable efforts have been made to understand fundamental problems in financial systems~\cite{bisias2012survey}, there is little work on system risk analysis in the networked-guarantee loans, except for preliminary work~\cite{meng2015credit}, where a positive correlation between the K-shell decomposition value of the network and default rates was reported. Readers are referred to~\cite{Yan2015Multi, Zhibin2015Rapidly} for more references on graph related applications.

\textbf{Visualization in financial systems} Visualization and visual analytics have been introduced to the financial sector, including transactions monitoring, price fluctuations, and complex decision-making~\cite{dumas2014financevis}. Animation is used to visually analyze large amounts of time-dependent data~\cite{archambault2016can,archambault2011animation}. The 3D treemap is introduced to monitor real-time stock market performance and to identify a particular stock that has produced unusual trading patterns~\cite{huang2009visualization}. The interactive exploratory tool is designed to help the casual decision-maker to quickly choose between various financial portfolios~\cite{rudolph2009finvis}. Coordinated specific keywords visualization within wire transactions are used to detect suspicious behaviors~\cite{chang2007wirevis}. The Self-Organizing Map (SOM), a neural-network-based visualization tool, is often used in financial risk visualization analysis for monitoring the occurrence of sovereign defaults in less developed countries~\cite{sarlin2011sovereign}, for the visual analysis of the evolution of currency crises by comparing clusters of crises between decades~\cite{sarlin2011clustering}, and for discovering imbalances in financial networks~\cite{sarlin2013chance}. Readers are referred to~\cite{lindskog2000modelling} for more references on financial visualization.
The visual analytic approach is also employed to analyze contagion in networks and in the simulation of contagion effects~\cite{von2015visual}. Motifs are employed to analyze and visualize the network~\cite{von2009system,klukas2006coordinated,maguire2013visual}. We are inspired by the various technologies and designed a visual interface for networked-guarantee loan risk management.
%-------------------------------------------------------------------------

\section{Risk Management and Visualization} %design reason, architecture
\begin{figure*}[!ht]\vspace{-15pt}
  \centering
  \includegraphics[width=1\linewidth]{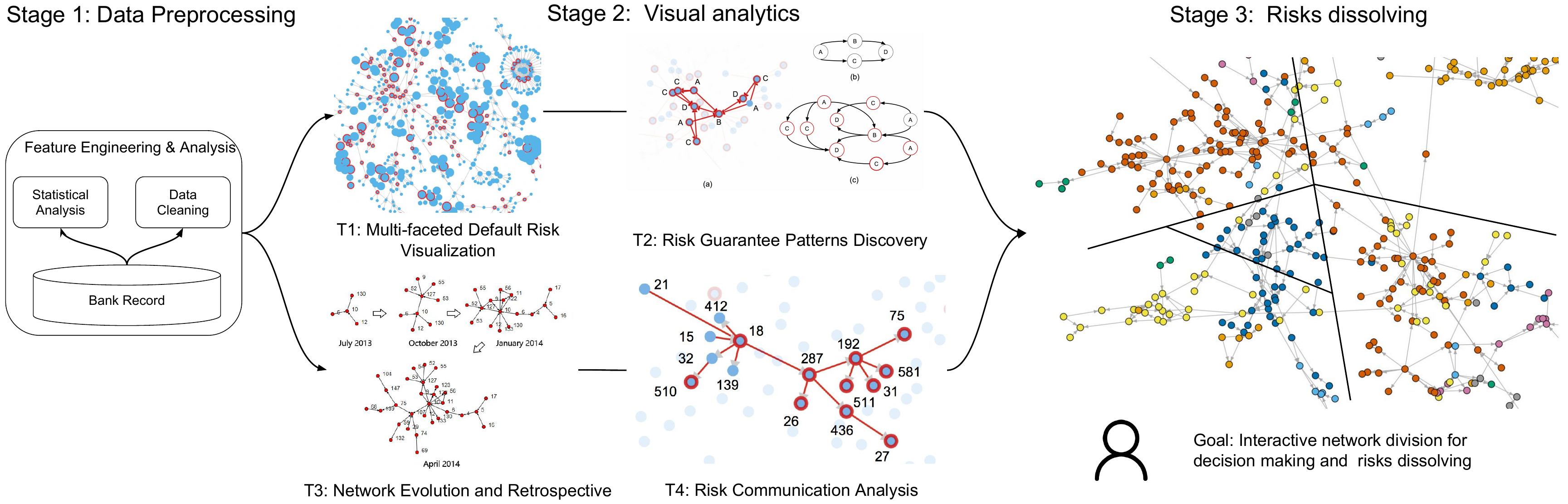} \vspace{-20pt}
  \caption{Overview of the system and tasks.}\label{overview}\vspace{-15pt}
\end{figure*}

%We consult with financial experts and set the ultimate goal of developing an interactive network division approach and tool for risk isolation and decision-making. Based on the goal, we consolidate four analysis tasks.  \autoref{overview} gives the overview of the system and tasks. In this section, we give a brief introduction before describing the detailed algorithm, strategy, and interactions. Specifically, the tasks include:

%possible malicious guarantee activity
We consult with financial experts and set the ultimate goal of developing an interactive tool for the government and banks to monitor
default spread status and provide insight for taking precautionary measures to prevent and dissolve systematic financial risk. Based on the goal, we consolidate four analysis tasks. The tasks include:

\begin{description}\vspace{-8pt}
  \item[Task1:] \textbf{Multi-faceted Default Risk Visualization}. The current loan credit rating system is based on the pure financial status of the individual borrower. The credit assessor can usually access the first layer of the guarantee chain, thus cannot trustfully evaluate the risks. It is necessary to carry out a systematic analysis of the enterprise to avoid inadequate risk assessment.\vspace{-8pt}
  \item[Task2:] \textbf{Risk Guarantee Patterns Discovery}. Fraud guarantee patterns may lead to default and diffusion. Identifying new high default patterns helps banking experts to single out and tackle the principal default problem. Visual analytics tools should be developed to thoroughly analyze the network.\vspace{-8pt}
  \item[Task3:]  \textbf{Network Evolution and Retrospective}. Understanding the network dynamic helps financial experts to understand how firms are connected together temporally. It requires visualizing the evolution of the guarantee network based on historical data.\vspace{-8pt}
  \item[Task4:] \textbf{Risk Communication Analysis}. Before a crisis occurs, forecasting the default diffusion path and monitoring the default spread status will help the government and banks to take precautionary measures, conduct research, and take effective measures to prevent and dissolve risks.\vspace{-8pt}
\end{description}

\autoref{overview} gives the workflow. In the data preprocessing stage, guarantee networks are constructed from the bank records. Then, the spatiotemporal information is utilized during the visual analytics stage.  In task 1, forecasted default risk and network related measurements are visualized to help to locate hotpot efficiently. In task 2, an interactive interface is designed to help the experts to explore and discover possible malicious loan frauds. In task 3, the evolution of the network provides insights of the past enterprises' activity and task 4 provides the possible default spread path in the future. In the risks dissolving stage, with the insights obtained from previous stage will help to divide the guarantee network so that no regional or systematic financial risks occur. We next describe the detailed algorithms, strategy, and interactions.

%competitive decision-making , such that no regional or systematic financial risks occur

\subsection{Default Risk Prediction and Visualization}

The loan records reveal that guarantee network and default rates are both increasing, and the network structures show a strong correlation with the defaults. We construct feature vectors consisting of hybrid information and employ the supervised learning approach to train the prediction model. In what follows, we discuss the hybrid features used in our model.

In order to build a highly representative feature that can reliably reflect the statistical relationships between the customers’ information and their repayment ability, we clean the data and construct the features as: (1) Basic Profile, the essential company registration information, which reflects the character, capital, collateral, capability, condition, and stability~\cite{meng2015credit}. We use business nature, registered capital, enterprise scale, employee numbers and other information to make up the corporation’s basic profile. Most banks require the company to update this basic information when the enterprise makes a loan application; we choose to use the latest information as the basic profile features of the loan. (2) Credit Behavior, historical behavior, e.g. credit history, default records, default amount, total loan amount and loan count, total loan frequency (if any), total default rates. This is calculated using all the loan records before the active loan contract. (3) Active Loan, the loan contract in its execution period. It contains active loan amount, active loan number, type of capital return and interest return, etc. (4) Network Structure, network features such as centralities are extracted. Note that, as discussed above, the basic profile may not be completely trustworthy, as the businesses may provide out-of-date or even false information to the bank. However, the guarantee network uses trustworthy information, as the bank can build the data from its own records.

The prediction of default for a customer’s loan guarantee can be modeled as a supervised learning problem. We choose to use logistic regression based on a gradient boosting tree to predict the risk for the reason that it is reportedly successful in many data science problems. Also, note that our task is to visualize the risk for different enterprises. We do not compare the prediction performance of various regression methods in this paper; these will be demonstrated in our future work.

In the XGboost, the tree ensemble model using $K$ additive function to prediction output can be represented as:\vspace{-5pt}
\begin{equation}\label{tree}
  \hat{{y}_{i}}=\sum_{k=1}^{K}f_{k}(X_{i})\vspace{-5pt}
\end{equation}

In Eq.~\ref{tree}, $f_{k}$ is the $k_{th}$ decision tree, $X_{i}$ is the training feature and $\hat{{y}_{i}}$ is prediction results. Finding parameters of the tree model is turned into minimizing the objective function problem and it can be trained in an additive manner~\cite{chen2016xgboost}. \vspace{-10pt}

%\begin{equation} \label{e1.1}
%A x = b
%\end{equation}

\begin{equation}
  L(\phi ) = \sum_{i}{l(\hat{y}_{i},y_{i})}+\sum_{k}\Omega (f_{k})\newline\vspace{-5pt} \ where \ \Omega (f) = \gamma T+\frac{1}{2}\lambda \left | \left | \omega  \right | \right | ^{2}
\end{equation}

where $\sum_{i}{l(\hat{y}_{i},y_{i})}$  is a training loss function that measures the difference between the prediction and the target; $\Omega (f)$ is a smoothing regularization term to avoid over-fitting.

\begin{figure}[tb!]\vspace{-2pt}
  \centering
  \includegraphics[width=1\linewidth]{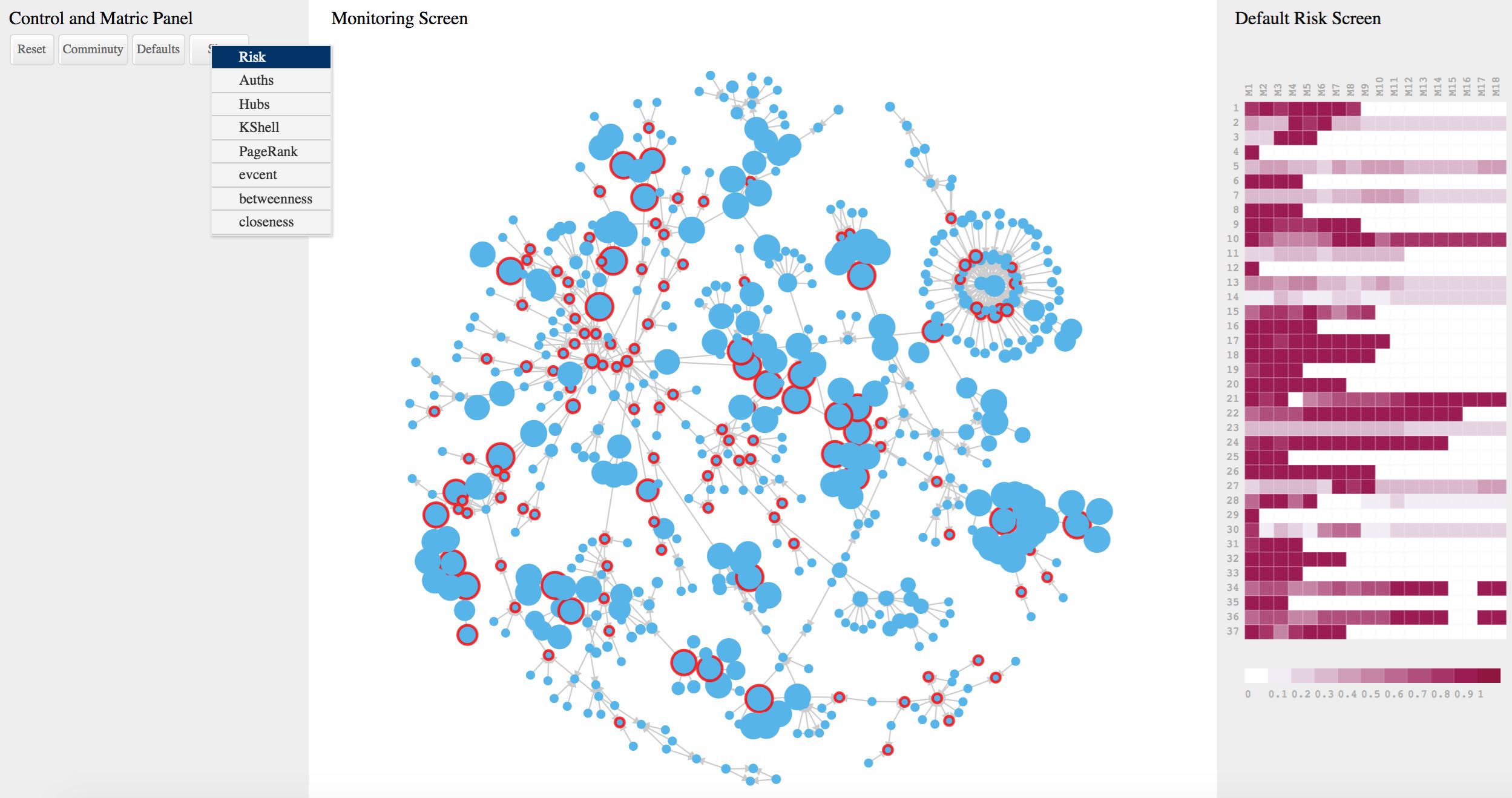}\vspace{-10pt}
  \caption{The interface for Visual Analytics for Enterprise Default Risk. We use a heatmap to code the rolling prediction risks over a month.}\label{heatmap}\vspace{-10pt}
\end{figure}

\begin{figure}[tb!]
  \centering
  \includegraphics[width=1\linewidth]{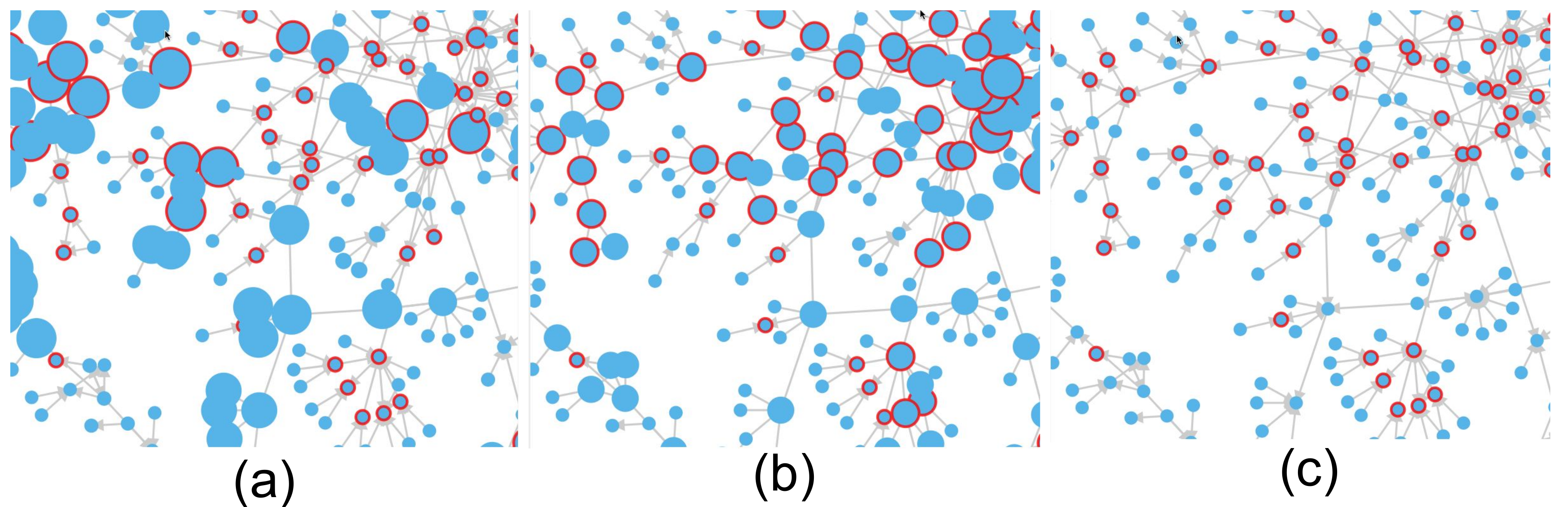}\vspace{-10pt}
  \caption{Visualization of the network with (a) the rolling prediction risk, (b) K-shell value, and (c) authority score.}\label{highlight}\vspace{-15pt}
\end{figure}

We design and implement a visual interface enabled to view the network with various multiple measurements. \autoref{heatmap} gives the interface, by which users can adjust the node size by the predicted default risk (proportional to the diameters of the sphere) and by the following network centrality measurements: Hub score and Authority score, K-Shell decomposition score, PageRank, Eigenvector centrality scores, Betweenness centrality, and Closeness centrality. \autoref{highlight} gives a part-visualization of a real guarantee network. In the graph, all defaulted enterprises are highlighted by red circles. Node size is proportional to predicted risk (a), K-shell value (b), and authority score (c). Through the interface, users can also observe the rolling prediction risk of an enterprise over a month and highlight it on the whole network by choosing it on the heatmap.

%The risk assessment framework is illustrated in \autoref{loanPeriod}. Firstly, the loan records (see Section 3.1 for details) are extracted from the data warehouse and stored in a customer data management (CDM) system. Then, five categories of features (basic profile, credit behavior, active loan, community behavior, and network structure) are extracted from the loan records in the given sliding window. These features serve as the model input.

\subsection{Risk Guarantee Patterns Discovery}
%We develop an interface enabled to automatically guarantee risk pattern detection and visualization -- the commonly recognized risk loan patterns, including the mutual guarantee, co-guarantee, and revolving guarantee, are highlighted on the network. Some guarantee circles are relatively clearly understood by banking experts.

\begin{figure}[tb!]
  \centering
  \includegraphics[width=1\linewidth]{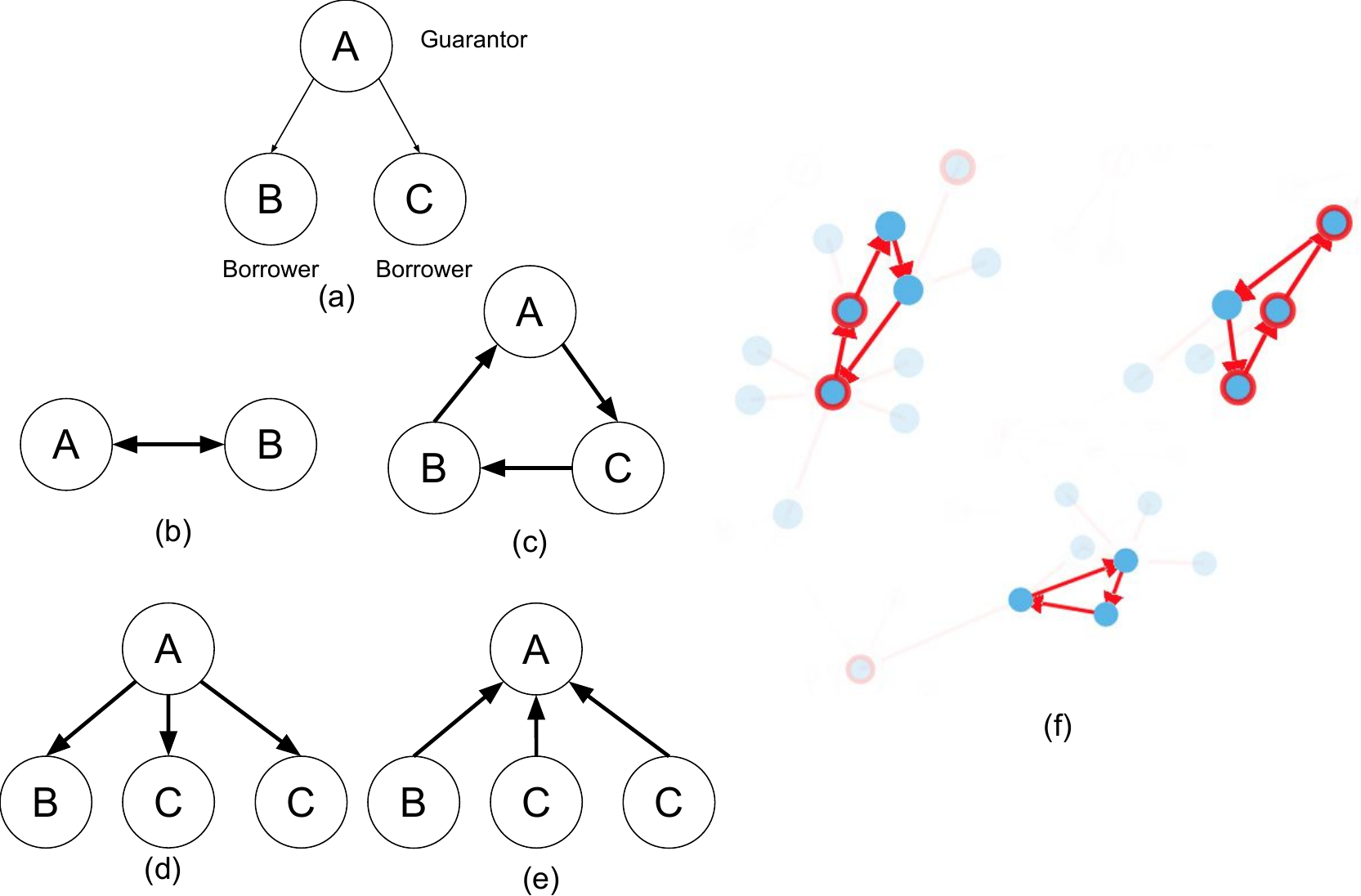}\vspace{-10pt}
  \caption{(a) guarantee network, where enterprise A (guarantor) guarantees B and C (borrowers) to get loans from the bank (lender). The (b--e) graphs are classic loan guarantee patterns, specifically: (b) mutual guarantee, (c) revolving guarantee, (d) star shape guarantee, (e) joint liability guarantee. (f) revolving guarantees detected from a real-world loan guarantee network.}\label{guaranteetypes}\vspace{-20pt}
\end{figure}

Empirical studies by bank risk control specialists suggest risk guarantee patterns, including mutual guarantee and revolving guarantee (see \autoref{guaranteetypes}). Such interactions are currently legal in China but in the banking industry, specialists in the bank risk control department only have SQL query capability to detect relatively simple guarantee patterns. Understanding more complicated risk guarantee patterns is difficult due to the tool’s limitation. An arbitrary guarantee pattern, which has a high default rate, can lie underneath the complex network structures. Thus, it is impossible exhaustively to compare all network patterns to determine whether it is in high default. In this work, we develop a visual analytics tool to help the experts explore, discover and further understand what has happened. We follow Ben Shneiderman’s mantra of information visualization, and the approach includes two steps: first, high default group detection; then risk guarantee pattern discovery.

\emph{High default group detection.} Recognizing high default groups narrows down the search scope of the risk guarantee relationship. Based on the conjecture that defaults tend to occur in clusters, we divide the whole network into several distinct sets by community detection. Theoretically, community structure in the graph is defined as the node sets that interact with each other internally more frequently than with those outside it. Identifying such substructures provides insight into understanding the structure of complex networks (both the functions and the topology affect each other).

We use a force-directed graph with colored communities and revised treemap interface to visualize the community detection results. The community label and default rates are displayed on the flat colored blocks. The treemap chart is used for navigation here; thus, the sum of the area does not necessarily need to be one. The larger blocks reveal the high default communities saliently.

\begin{figure}[tb!]
  \centering
  \includegraphics[width=1\linewidth]{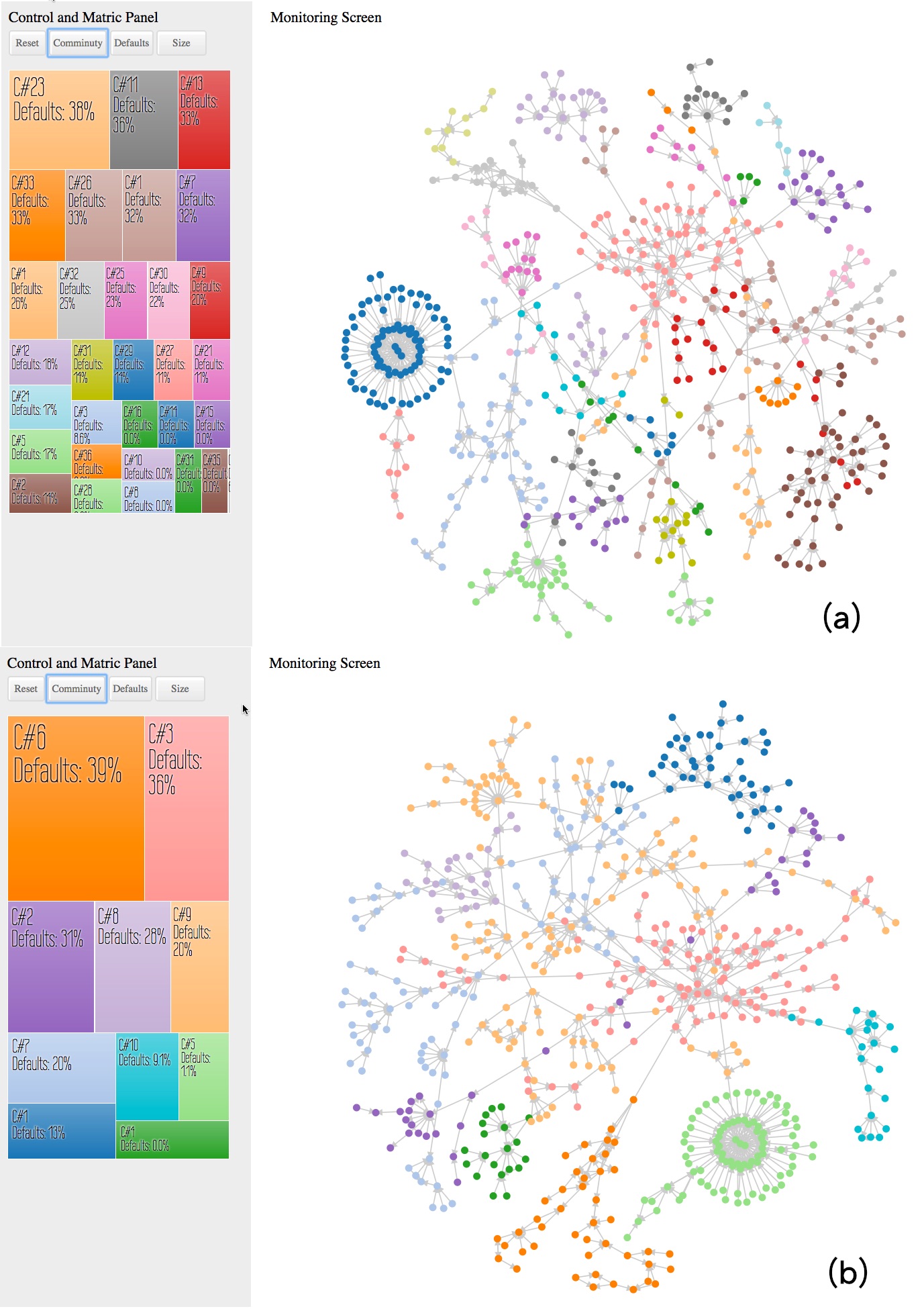}\vspace{-10pt}
  \caption{Defaults occur in clusters and we interactively edit the clusters. (a) 30 communities generated by a random walks algorithm; (b) 10 communities after interactive editing. The ratios for defaulting firms are labeled separately on the left-hand side treemaps. }\label{wc}\vspace{-20pt}
\end{figure}

\autoref{wc} (a) shows the results on a typical independent subgraph that we constructed from bank loan records. The communities are marked using a separate color background and the average default rates are labeled. There are 36 communities, of which defaults occur in 27, with an average 38\% to 8.6\% default rate, all other 9 communities have no defaults. We adopt the random walk algorithm~\cite{rosvall2008maps}. A similar phenomenon is observed on random walks, edge betweenness, and spineless community.

However, the evaluation of community detection is still an open question~\cite{lancichinetti2008benchmark}. As the community detection algorithm only considers the link information and neglects the node attribute information, the partition may not be consonant to the actual conditions. The basic rule for community detection is to minimize the number of links between communities, and this uses pure network structure information. In financial practice, each node in the network comes with rich information, such as enterprise sectors, changes in deposits, assets, loan amount, etc. It would be unreliable to discard such attributes when dividing the network. By interaction, we enable the users to edit the communities into coherent ones by referring to the relevant financial metric. We allow users to interactively perform the following manipulation actions.

\emph{Interactive community editing}. We enable users to explore the financial information and interactively edit the communities by merging strongly associated communities, to reassign the community labels for the structural hole spanners (a key role in the information diffusion) ~\cite{burt2007secondhand}, or to split a community into several distinct smaller groups. The generated subgraphs are noted as groups of interest (GOI), in which the high-risk guarantee pattern is often hidden.

\begin{figure}[tb!]
  \centering
  \includegraphics[width=1\linewidth]{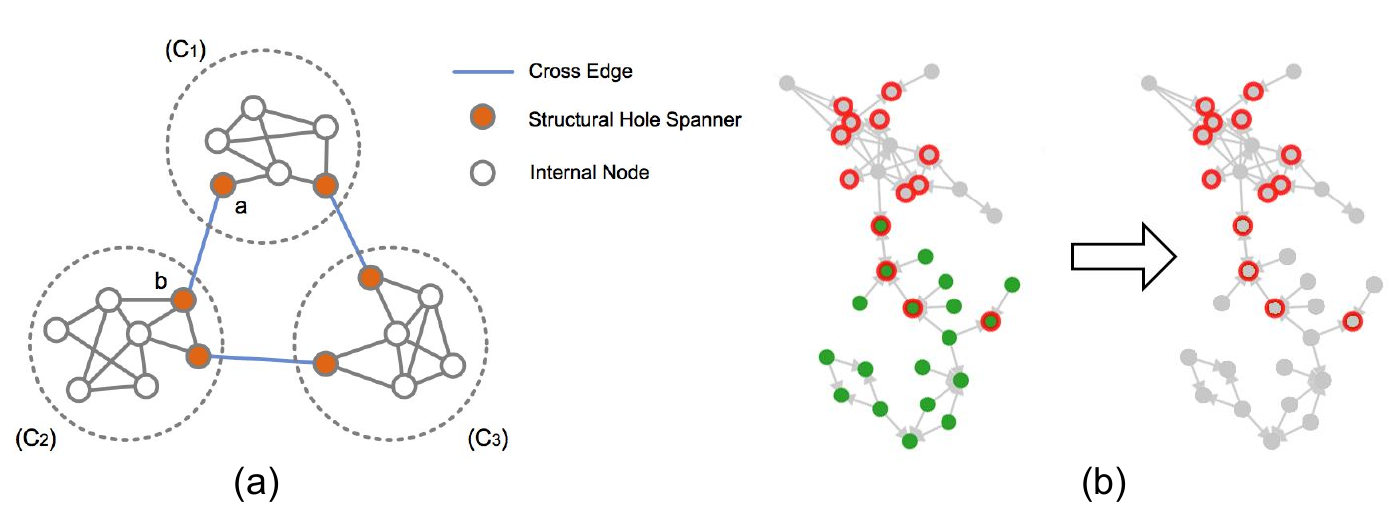}\vspace{-10pt}
  \caption{(a) Structural hole spanner illustration example, reproduced from~\cite{he2016joint}; the structural hole spanners are editable for merging or reassigning actions. (b) Example of merging two communities on the structural hole spanner.}\label{hole}\vspace{-20pt}
\end{figure}

\textbf{Reassign}. The reassign operation allows the user to change the community labels of the structure hole spanner. The structure hole spanner is the bridge node that connects different communities in a network. \autoref{hole} is reproduced from~\cite{he2016joint}; it illustrates a network with three communities and six structural hole spanners. Empirical study suggests that individuals would benefit from filling the ``holes'' (an alternate name for the structural hole spanners) between people or groups that are otherwise unconnected~\cite{burt2004structural}. We observed high default in structure hole spanners with their neighboring internal nodes. We enable the users to investigate the financial metrics and reassign the community labels of the structure hole spanners. With the interface, he/she first double-clicks the ``tile''  on the treemap highlighting all the connected communities. Single-clicking the structure hole spanner node can reassign it to the opposite community.

\textbf{Merge}. Naive community detection divides a graph based purely on links in the graph, it may generate many communities where some of them share a common sector category or similar network structures. Merging the communities referring to a specific financial metric can produce medium-sized and more tractable subgraphs. With the interface, he/she first double-clicks one ``tile'' on the treemap highlighting all the connected communities and then double-clicking the structure hole spanner node to merge the two communities.

\textbf{Split}. When the default is unevenly distributed, we need to split the community and cut off the stable parts to reduce the next motif related computation complexity. With the interface, he/she first double-clicks one ``tile'' on the treemap highlighting the connected communities and then double-click the edge making the two opposite parts of the subgraph be split into two communities.

The interface also has a financial radar view to encode the key financial infromation. The key indices include:
\emph{Defaults}, historic default behavior; \emph{LA/RC} the ratio of loan amount to registered capital. It would be more insightful to use the ratio of loan amount to enterprise net assets; however, the latter information is not always available, so we use registered capital instead. \emph{Deposit loss} the rapid decrease of deposit and shorting of money may imply business out of the situation. \emph{Sector} the enterprise sector related to the macroeconomic conditions and is an important clue when editing communities. \emph{GA/RC} the ratio of guarantee amounts to registered capital. The ratio of guarantee amount to enterprise net assets is a crucial factor for the stability of the financial system. Also because of lacking information transparency, we use registered capital instead. \emph{Credit rating} is the review rating of bank experts; this is also a key clue when editing communities.

\emph{Risk Guarantee Pattern Discovery and Visualization.} The guarantee patterns that are prone to default may exist underneath the GOIs. A complex guarantee network is always connected by several smaller subgraphs bridged by the structural hole spanners. The subgraphs inside the communities may reveal certain risk patterns; even a fraud pattern. The motifs are the most basic building blocks for a graph and they may reflect functional properties. In this work, we obtain a set of motifs by first detecting motifs from the GOI. The motifs are ranked by their default rates (Eq.~(\ref{rglp})). High default rate motifs are noted as a pattern of interest (POI); these may need to be investigated by banking experts as a priority.

\begin{equation}\label{rglp}
priority = (\frac{\sum default\_node\_number(m))}{\sum node\_number(m)}) \vspace{-5pt}
\end{equation}
where $m$ is a motif. All motifs are possible risk guarantee patterns.

However, it is still computationally challenging to obtain all POIs using the approach above for the following reasons. Firstly, motif structures increase rapidly with an increase in node number; for example, a four-node motif gives rise to more than 3000 possibilities. It is therefore impossible to enumerate all the motif structures. Secondly, motif matching is exhaustively searched from the query graph into the large network, and it presents a subgraph isomorphism problem. It still takes too much time for motifs with more nodes to be matched on the network. With the interface, we enable interactive motif editing. Users can refer to the  financial radar view of adjacent nodes and add new nodes to the motifs to generate a more complex POI without exhaustively compute all possibilities.

\subsection{Network Evolution and Retrospective}

Network evolution over time is observed from the guarantee network. The topology of the network keeps changing: some nodes are connected to the network or removed from it; some communities are connected together through the guarantee of the structural hole spanner. Like many other real networks, competitive decision-making is taking place in the guarantee network: When a firm lacks the security to obtain a loan from a bank, it may resort to a guarantee corporation or third-party firms. To some extent, the new guarantors may improve the overall rationality of the system, but may also induce an unstable factor as the network becomes even more complex. Understanding the network dynamic helps financial experts to understand how the firms are connected together temporally.

\begin{figure}[htb!]
\begin{center}
\includegraphics[width=1\linewidth]{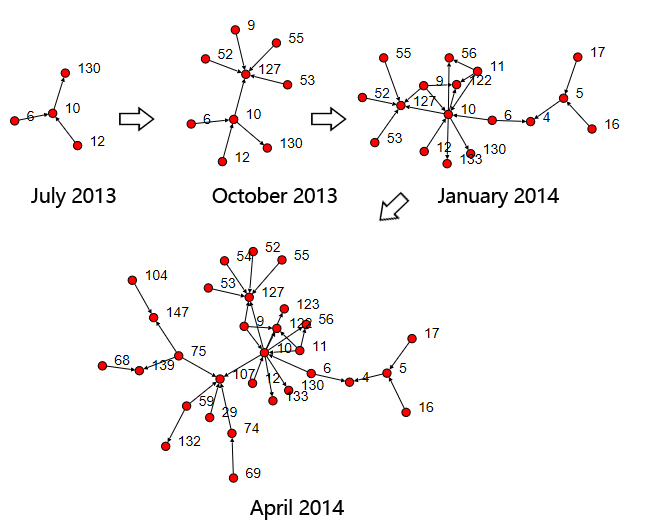}\vspace{-10pt}
\caption{The guarantee network keeps evolving from July 2013 to April 2014. The numbers in the graph are node ID}\label{evolve}\vspace{-30pt}
\end{center}
\end{figure}

Animation is employed to visualize the evolution of a guarantee network. With the interface, users can drag the time bar to backtrack how the network has evolved over time. They can hover the cursor over a node to view the company’s financial information. This will help the financial experts to understand what has happened historically. \autoref{evolve} gives an example of a real network evolved from July 2013 to April 2014. By combining enterprises financial information of different time, financial experts would be able to make the analysis.
\begin{figure}[tb!]
  \centering
  \includegraphics[width=1\linewidth]{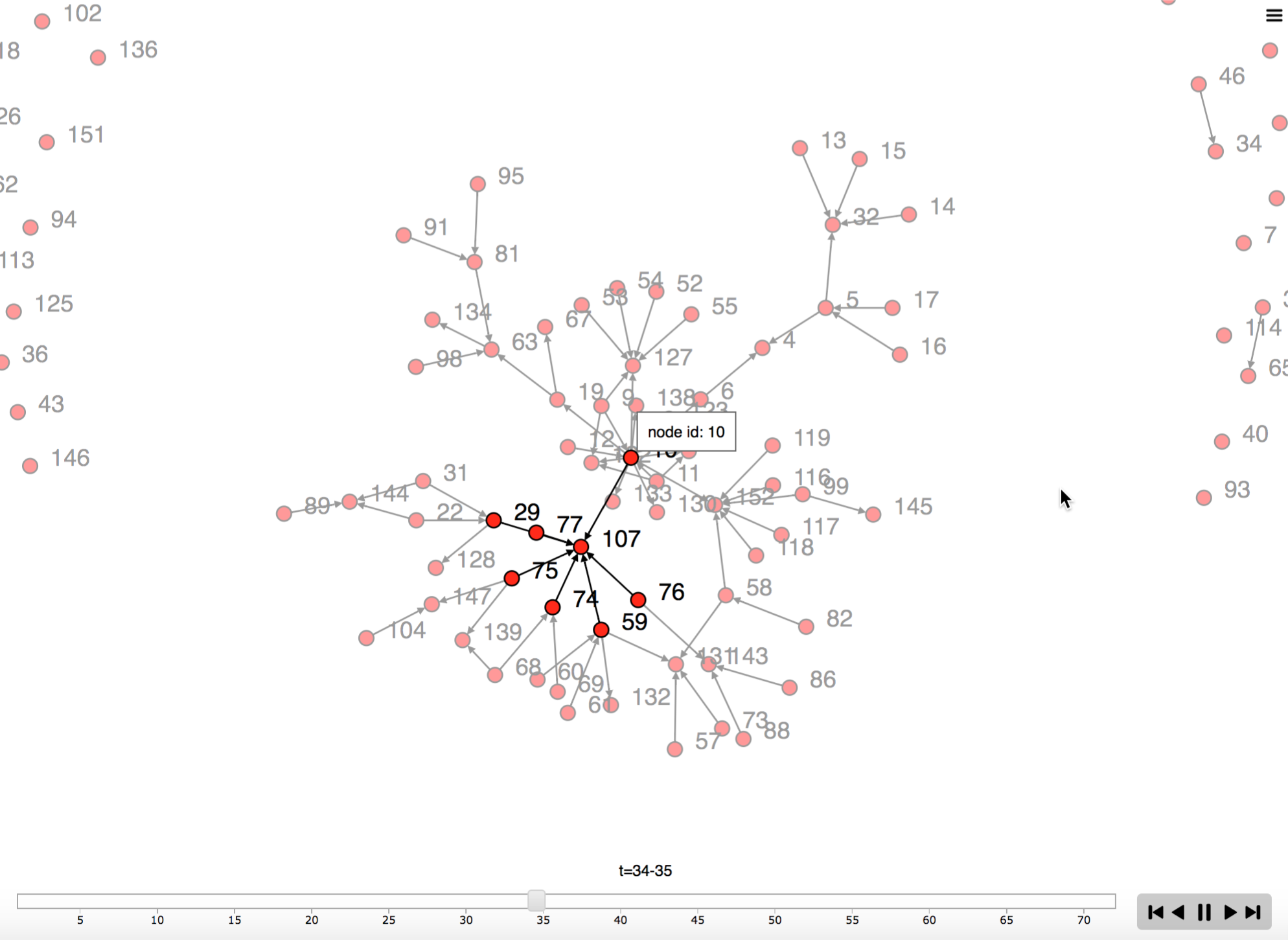}\vspace{-10pt}
  \caption{Visual analytics interface for evolving loan guarantees. The numbers in the graph are node ID}\label{circle}\vspace{-10pt}
\end{figure}

\subsection{ Risk Communication Analysis}
As a new phenomenon, the understanding of the systematic risk of the networked-guarantee loan is still insufficient. Sophisticated guarantee relationships tend to cause credit granted by multiple lenders and excessive credit. In the loan guarantee, a guarantor takes on the debt obligation if the borrower defaults; therefore, if the guarantee cannot be paid back to the bank, it may resort to its guarantors. In this case, the default may propagate throughout the network, like a virus. The default contagion increases both the possibility of the occurrence of risks and the transmission of risks. Especially in a period of economic downturn, some enterprises will face operational difficulties and the financial crisis will have a domino effect: the default phenomenon may spread rapidly in the network, and this could make a large number of enterprises fall into an unfavorable situation. The government and the banks always wish to monitor the default spread status and understand the complexity of the current issue of risks so that they can take precautionary measures, conduct research, and dissolve the risks to ensure that no regional or systematic financial risk occurs.

\begin{figure}[ht]
\begin{center}
\includegraphics[width=1\linewidth]{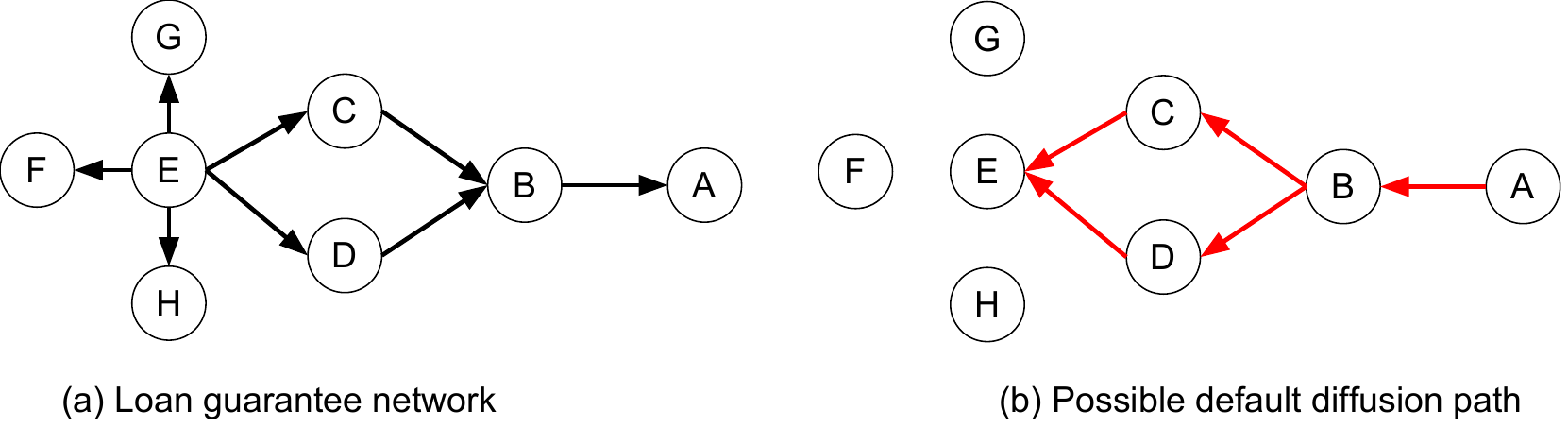}\vspace{-10pt}
\caption{Default path for a real network. The characters represent different enterprises. }\label{pathdemo}\vspace{-20pt}
\end{center}
\end{figure}

Based on relevant knowledge and experience, we develop a visual analytics tool to aid the default path discovery by visualization. A principle of the default diffusion can be described, as the vulnerable nodes are the guarantors. \autoref{pathdemo} gives a diffusion path illustration. (a) is a guaranteed network with eight nodes, where node E provides a guarantee to five adjacent nodes and C, D provide a guarantee to B and then to A; (b) is the possible diffusion path: the default of node A may lead to B, C, D, and even E defaulting. It is noted that nodes G, F, and H are not connected with node E, and therefore the default of E will not affect the repayment status of G, F, or H.

\begin{figure*}[ht!]
\begin{center}
\includegraphics[width=1\linewidth]{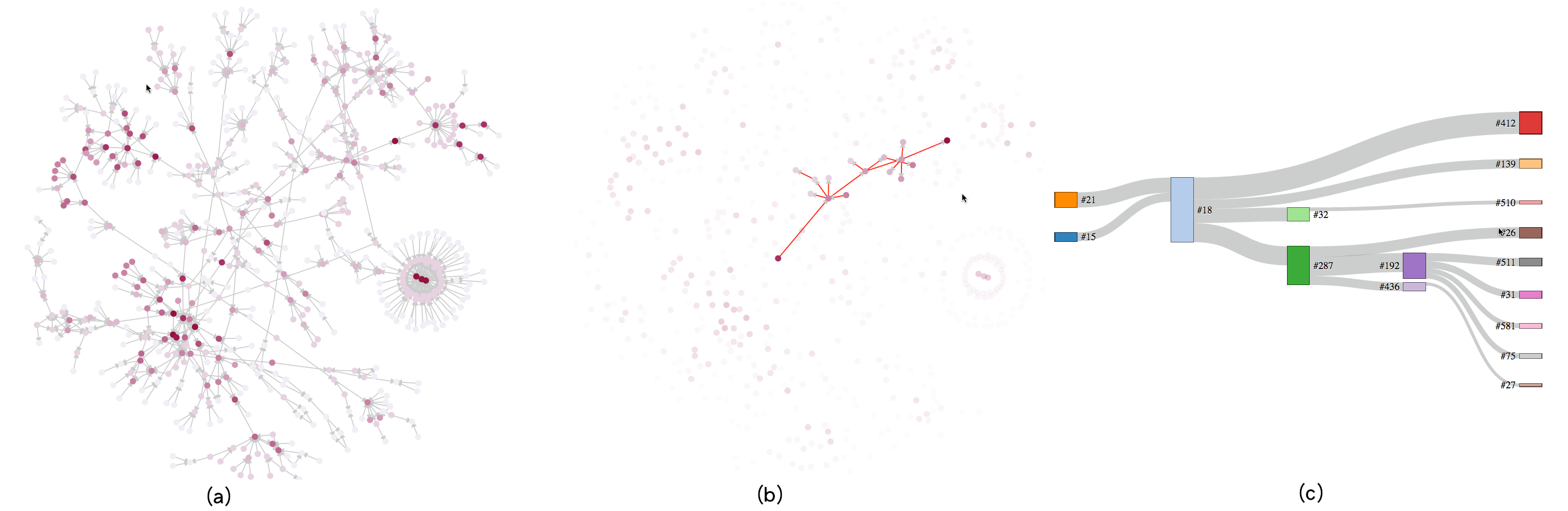}\vspace{-10pt}
\caption{One real diffusion path and the corresponding Sankey diagram. }\label{sankey}\vspace{-20pt}
\end{center}
\end{figure*}

In practice, there may be multiple possible propagation paths, as each node can serve as a guarantor or get guaranteed. It is difficult to outline the main propagation path from the entire graph. We make the following assumption: the node on multiple propagation paths is the key to prevent large-scale default diffusion and thus should be highlighted. We compute all the propagation paths, count occurrences, and highlight the node on the network. We use color to illustrate the \emph{propagation risk importance}.

We design the visual analytics tool, which enables financial experts to take into account several factors on the judgment of defaults. These factors include the financial information on the corporation and the guarantee contract amount information. The former information is plainly listed when the user hovers the mouse pointer over the node, while a Sankey diagram is used to represent the guarantee flow. The widths of the Sankey diagram bands are directly proportional to the guarantee amount.

\autoref{sankey} (a) gives results on a real guarantee network, when we choose one node—for example, node 32. The whole potential propagation path is highlighted in (b), while (c) is the corresponding Sankey diagram. It can be seen that upstream companies usually provide more in guarantees than they receive. For example, node 18 provides much more guarantee than it receives. The imbalance between the guarantee amount and the collateral amount provides a clue for credit line assessment. The real situation is even more complex. The default may be diffused like a viral infection, and the virus must identify and bind to its receptor (guarantor). As mentioned earlier, each enterprise has more than 3000 financial entries, and it is therefore difficult to quantify each enterprise’s ability to resist infection. We enable users to look up multiple financial statuses and cut off the propagation path. We also note that the propagation model provides more insights to end users, and we plan to perform an in-depth study of the topic and provide a simulation interface in the future.

\section{Case Study}

\begin{figure}[tb!]
  \centering
  \includegraphics[width=1\linewidth]{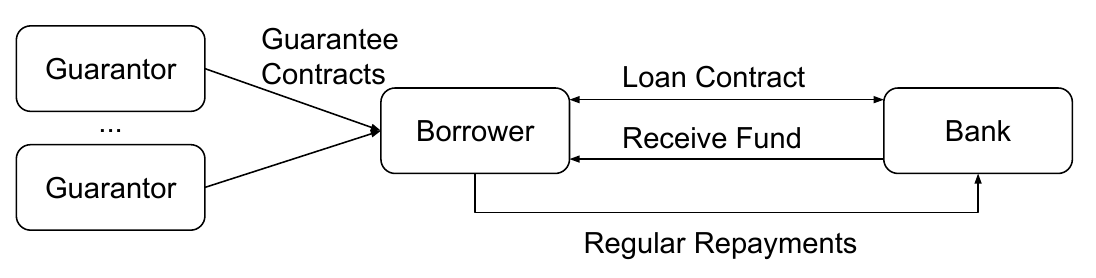}\vspace{-10pt}
  \caption{Loan guarantee process. The borrower wishing to get a loan from a bank first needs to sign loan guarantee contracts with guarantors before signing a loan contract. After the company receives its loan from the bank, it repays the loan by installments.}\label{glp}\vspace{-8pt}
\end{figure}

We first introduce the loan process, data exploration and then describe the experiments. As \autoref{glp} shows, there is often more than one guarantor per loan transaction, and there may be several loan transactions for a single guarantor in a period. Once the loan is approved, the business can usually obtain the full size of the loan immediately, and starts to repay the bank regularly by an installment plan until the end of the loan contract. The banks need to collect as much fine-grained information as possible concerning the repayment ability of the enterprise. The information falls into four categories: transaction information; customer information; asset information such as mortgage status; and history of loan approval records, etc.

We collect loan records spanning ten years from our cooperated commercial bank and construct the guaranteed network. The names of the customers in the records are encrypted and replaced by an ID.
%There are 11,000 loan customers, which span 60,948 mutual guarantee relationships with 5,911 defaults.

%\textbf{Overall statistics} There are 11,000 loan customers, which span 60,948 mutual guarantee relationships derived from 36,618 loan contracts. There have been 5,911 defaults during the past ten years, out of the total 87,307 repayments. The overall default rate for the number of contracts is 6.77\%.

\subsection{Multi-faceted Default Risk Visualization}\vspace{-2pt}

We propose a multi-faceted default risk visualization interface (see \autoref{heatmap}) and it includes forecasting default risk, centrality measurements (Authority score, hub score, K-shell, PageRank, Event, Betweeness, and, Closeness). We next explain them separately.

\emph{Default risk prediction}. As illustrated in Section 3.1, a hybrid representation and gradient boosting tree based approach is employed to predict the default risk. In the following experiments, we define Node-wise (NW) feature as the vector composed of basic profile, credit behavior, active loan information; define Network (N) feature as only network structure features; define Community Behavior (CB) feature as loan history behavior associated with graph community; define Hybrid (H) feature consists of both node-wise feature, network feature, and community behavior feature.

\begin{figure}[tb!]\vspace{-8pt}
  \centering
  \includegraphics[width=0.9\linewidth]{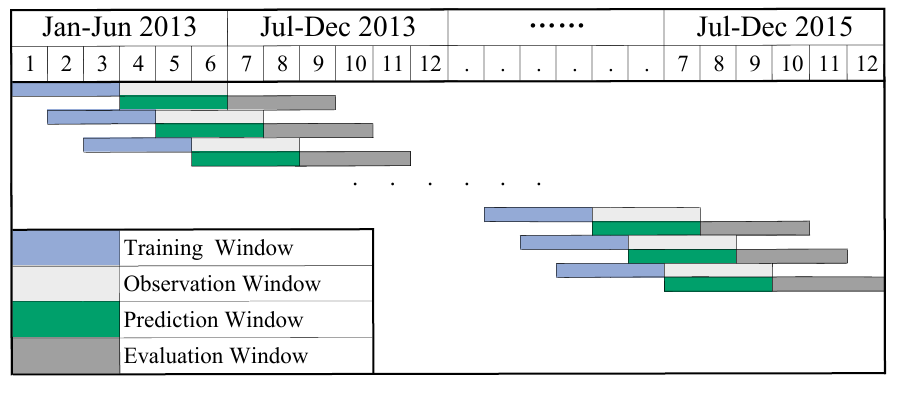}
  \caption{Illustration for the rolling sliding windows protocol. Features are extracted in the training window, and the corresponding outcome default label is collected in the observation window. Then the features and default outcome are used to train the model. The trained model is used by collecting the input features during the prediction window and verifying its performance when we reach the end of the evaluation window.}\label{window}\vspace{-8pt}
\end{figure}

Besides, we choose to employ a three-month sliding window setting for training, observation, prediction, and evaluation. The reasons are two-folds: (1) Prediction shall be adapted to a dynamic setting with a regularly updated forecasting results. In fact, using sliding window is a typical way for rolling prediction as commonly adopted in event prediction practices. (2) The business often runs on a quarterly basis, which can also be observed from the record that the default happens intensively at each end of quarter. Thus from a business demand perspective, it would be helpful to know the borrowers who may be default on a quarterly basis. As \autoref{window} shows, in the training stage, for all customers who obtain bank loans from 2013 Q1 (first quarter of 2013), the features are extracted in that period; the repayment status in 2013 Q2 are the labels to train the model. In the testing stage, we use the trained model to predict the customers who obtain loans between 2013 Q2 and use the real repayment status from 2013 Q3 to evaluate the performance when reaching the end of September 2013.

\begin{table}%\vspace{-10pt}
\tiny
\centering
\caption{AUC of forecasting models} \label{metrics}
\begin{tabular}{l|c|c|c|c} \hline
\hline
Period & NW &NW,CB & NW, N & H \\ \hline
\hline
2013 Q3  & 0.910 	& 0.924 	 &  0.917 &	 0.925  \\ \hline
2013 Q4  & 0.905 	& 0.926 	 &  0.920 &	 0.931  \\ \hline
2014 Q1  & 0.901 	& 0.929 	 &  0.923 &	 0.930  \\ \hline
2014 Q2  & 0.907 	& 0.931 	 &  0.928 &	 0.933  \\ \hline
2014 Q3  & 0.908 	& 0.935 	 &  0.933 &	 0.937  \\ \hline
2014 Q4  & 0.910 	& 0.933 	 &  0.939 &	 0.941  \\ \hline
2015 Q1  & 0.908 	& 0.937 	 &  0.946 &	 0.946  \\ \hline
2015 Q2  & 0.902 	& 0.938 	 &  0.942 &	 0.945  \\ \hline
2015 Q3  & 0.911 	& 0.935 	 &  0.946 &	 0.952  \\ \hline
2015 Q4  & 0.907 	& 0.935 	 &  0.954 &	 0.959  \\
\hline
\hline\end{tabular}\vspace{-5pt}
\end{table}

\begin{figure}[tb!] %\vspace{-10pt}
  \centering
  \includegraphics[ width=.8\linewidth]{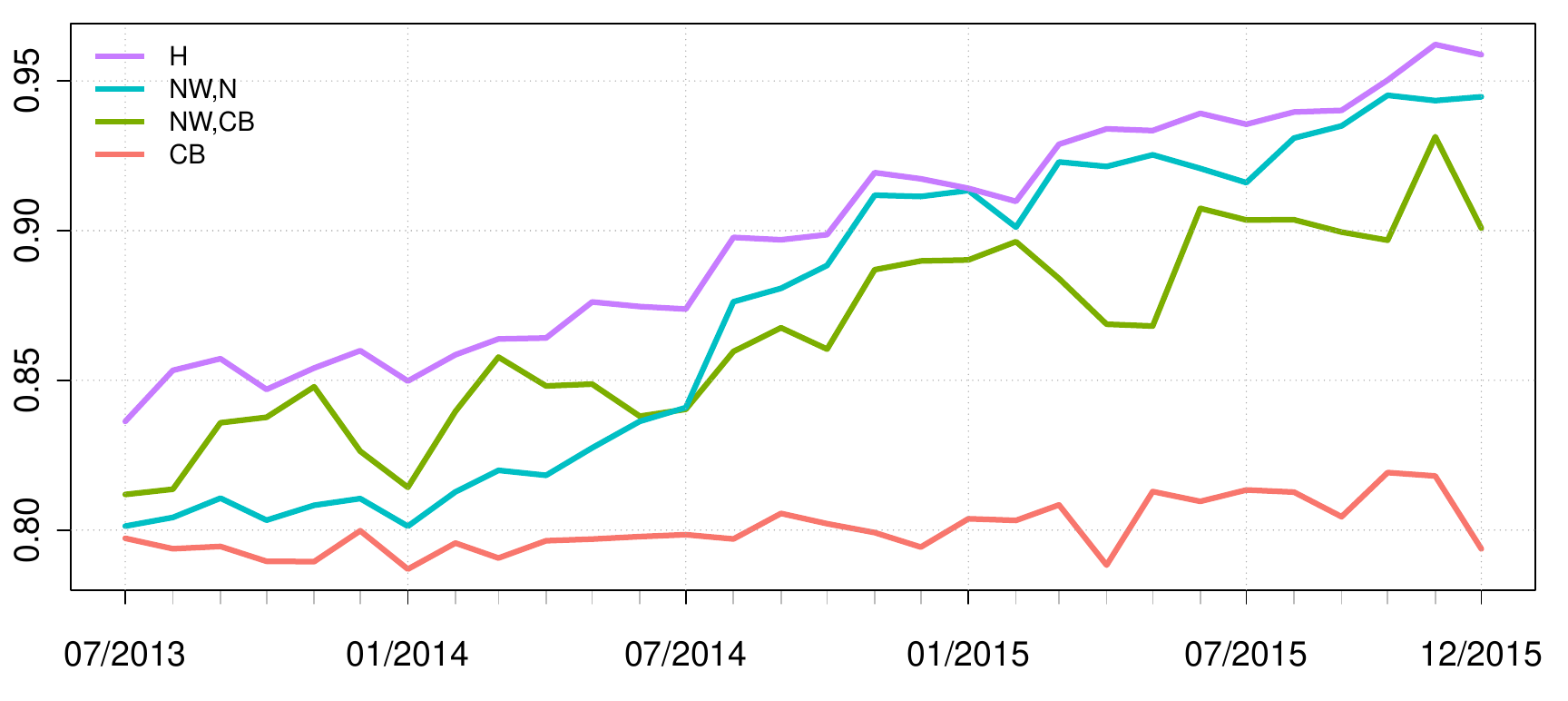}\vspace{-5pt}
  \caption{Recall of forecasting models using different feature representation over time. Refer to Section 4.1 for the abbreviations.}\label{plotRecall}\vspace{-10pt}
\end{figure}

We perform risk predictions using the proposed hybrid representation via an ablation test. The AUC (Area under Cure) of the models with different sliding windows are listed in Table~\ref{metrics}. As expected, the models using the hybrid feature always outperform other models with naive node-wise feature. It is worth noting that before 2014 Q4, the node-wise and community behavior feature (NW,CB) performs better than node-wise and network (NW,N) feature yet the latter outperforms since 2014 Q4. The recall curves in Figure~\ref{plotRecall} also reveal such a phenomenon, which perhaps is attributed to the increase of guarantee network complexity over time.

%It is known that it has to make a trad off between precision and recall values for a predication problem. In our problem, recall is more important as banks would like to aggressively identify all potential risk.

\begin{figure}[tb!]\vspace{-10pt}
  \centering
  \includegraphics[width=.8\linewidth]{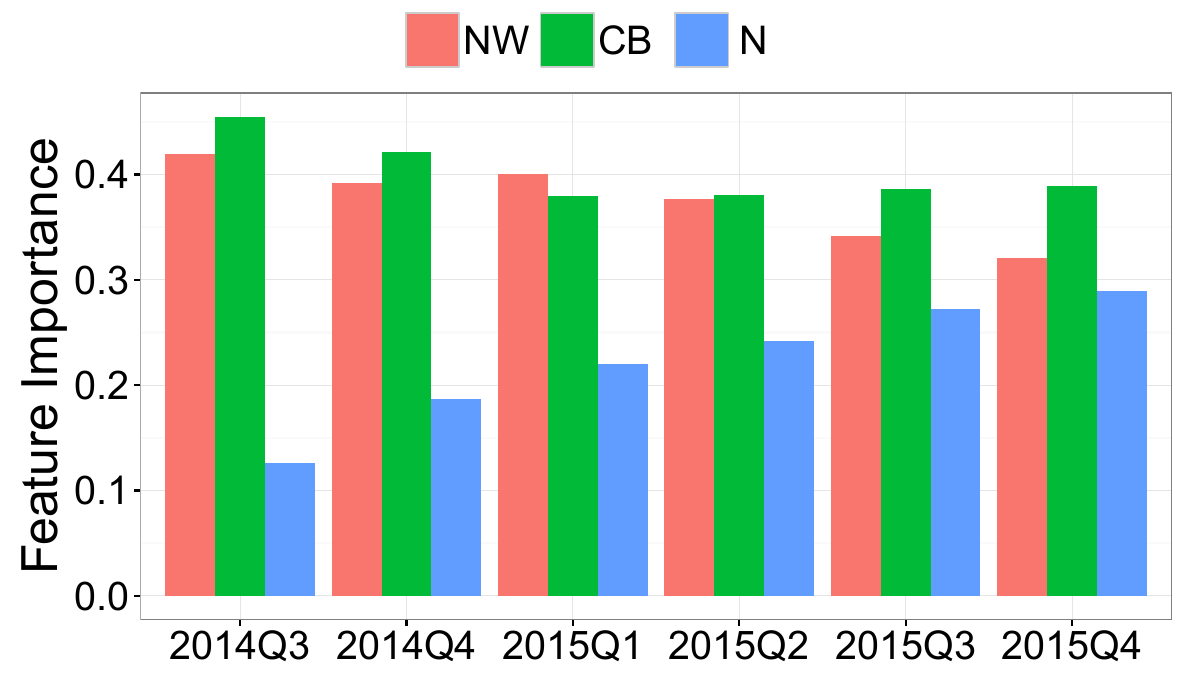}\vspace{-10pt}
  \caption{Feature importance score from 2014Q3 to 2015Q4. Refer to Section 4.1 for the abbreviations.}\label{fiaf}\vspace{-10pt}
\end{figure}

We also compare the prediction importance of node-wise, network, community behavior and our hybrid feature representation. By counting the times each feature is split to a branch of a decision tree in XGBoost regression, we can obtain relative importance of the features. As Figure~\ref{fiaf} shows, node-wise feature, community behavior and network feature take opposite trends over time. Initially, node-wise and community behavior features share similar weights and four times more than network features; With the network structure more and more complex, the network feature importance are increased and even account for nearly one-third importance at 2015Q4. This is consistent with the statistics observation that as the guarantee relationships becomes more complex over time, the network centrality related features become more important. Moreover, since node-wise feature only assumes customers are independent, it has weak discriminations when the enterprise are involved in a complex network.

\begin{figure*}[tb!]\vspace{-10pt}
  \centering
  \includegraphics[width=1\linewidth]{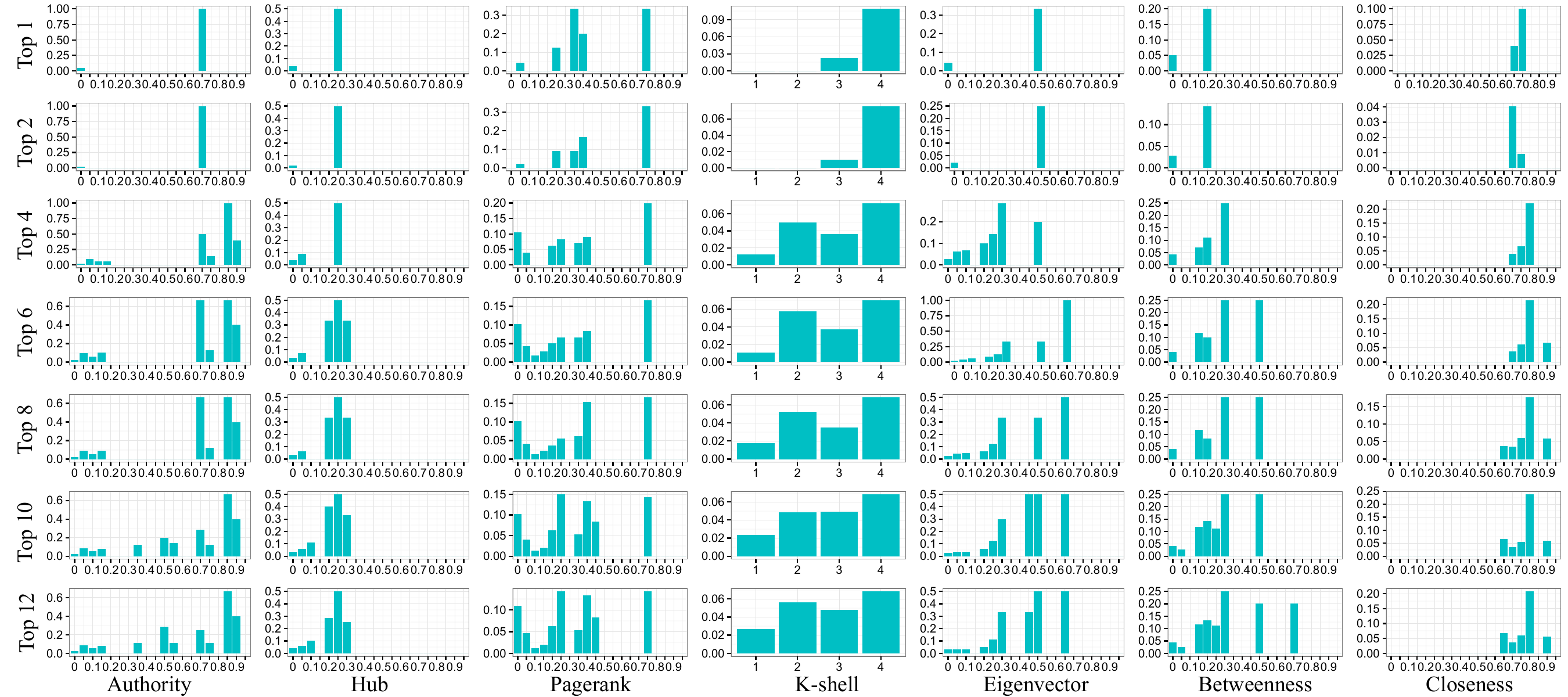}\vspace{-10pt}
  \caption{Overdue rates for different graph metric values. From left to right, each column is for a kind of graph metric, namely Authority score, Hub score, PageRank value, K-shell value, Eigenvector centrality, Betweenness centrality, Closeness centrality; From the top down, each row is the most complex independent subgraph.}\label{centrailty}\vspace{-10pt}
\end{figure*}

\emph{Centrality measurements}. We now report some observations derived from the data. Centrality indicators are helpful to identify the relative importance of nodes in the network. \autoref{centrailty} gives the histogram of several of the most complex subgraphs on how the defaults are distributed with different centrality indicator values. It is noted that defaults occur more frequently on nodes with large authority values and small hub values. This is consistent with intuition -- if an enterprise works as a hub and backs a large number of other corporations, it can be supposed that it is relatively stable and operates in good condition. Conversely, if an enterprise works as an authority and accepts guarantees from many other corporations, this is an indication that it lacks funding security and is at a higher risk of trouble. The statistics signal to the lender that it should watch the status of the ``authority'' high nodes in the guarantee network. Although the underlying assumption of PageRank is quite like the authority score, we did not observe a similar correlation between the values and the default rates (see \autoref{centrailty}).

However, it is difficult to reliably quantify the correlation of graph centrality indicators with enterprise defaults, in this case, interactive analytics tools provides the possibility to fuse the financial expert domain knowledge with the data-driven indicators. In the multi-faceted risk visualization interface (see \autoref{heatmap}) , different risks are highlighted by various diameter spheres and the users are able to explore enterprises from different point of views. This will help they make a better decision in the following analysis tasks.
\vspace{-5pt}

\subsection{High Default Group Detection}\vspace{-2pt}
High default group detection can reduce analysis scope and thus further help risk pattern discovery and it usually includes automatically community detection and interactive community editing. The experiments is performed on a independent guarantee network with 116 nodes. It is first automatically divided it into 36 communities. The statistics are given in \autoref{c36}.
\begin{table}[tbp]
\tiny
\begin{tabular}{|l|l|l|l|l|l|l|l|l|l|l|}
\hline
Community ID & 1    & 2    & 3    & 4    & 5    & 32   & 33    & 34 & 35   \\ \hline
Firms   & 44   & 42   & 35   & 19   & 29   & 4    & 3     & 4  & 4      \\ \hline
Defaults    & 14   & 6    & 3    & 5    & 5    & 1    & 1     & 0  & 0      \\ \hline
Ratio for default firms & 32\% & 14\% & 9\%  & 26\% & 17\% & 25\% & 33\%  & 0  & 0      \\ \hline
Ratio for default amount  & 68\% & 37\% & 4\%  & 92\% & 83\% & 72\% & 100\% & 0  & 0     \\ \hline
Structural hole spanner   & 7    & 3    & 5    & 2    & 2    & 1    & 1     & 1  & 1      \\ \hline
Neighbour communities    & 5    & 3    & 4    & 3    & 2    & 1    & 1     & 1  & 1      \\ \hline
Total loan amount    & 1071 & 518  & 1503 & 292  & 1282 & 18   & 48    & 57 & 105  \\ \hline
Total default amount    & 733  & 190  & 62   & 270  & 1065 & 13   & 48    & 0  & 0      \\ \hline
\end{tabular}
\caption{Statistics for communities generated by the random walk
community detection algorithm~\cite{rosvall2008maps}.}\vspace{-23pt}
\label{c36}
\end{table}

\begin{figure*}[tb!]
  \centering
  \includegraphics[width=1\linewidth]{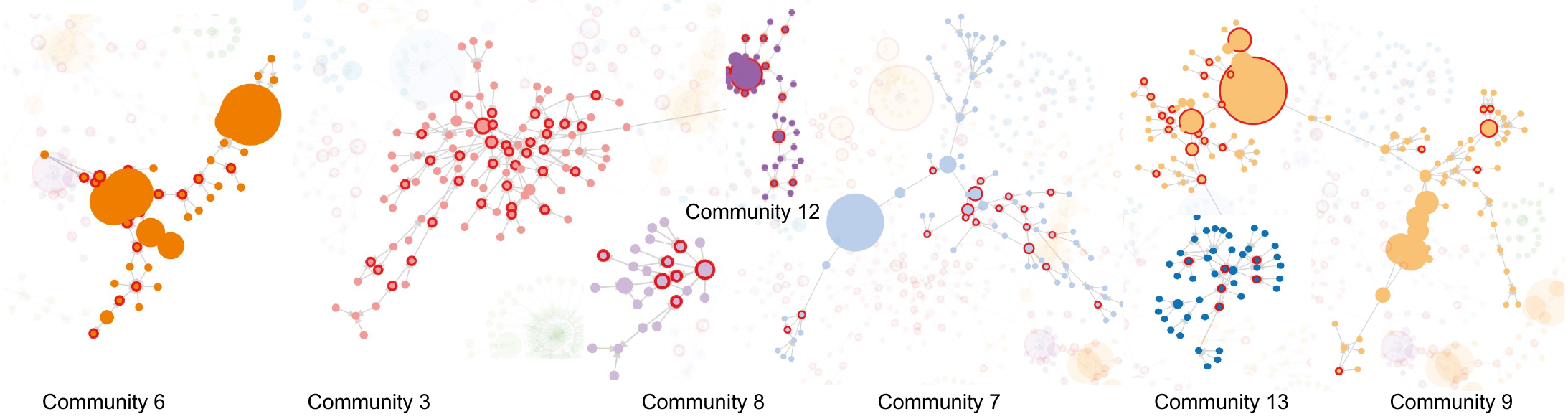}\vspace{-10pt}
  \caption{High default groups after interactive editing.}\label{merged}%\vspace{-10pt}
\end{figure*}

We edit the community following basic guidelines: (1) consider default status, loan amount, and other financial statistics comprehensively; (2) small communities can be either merged with large neighboring communities or pruned. For example, communities 35 and 34 both have four nodes and these firms never default. There is a low possibility that they will become high default groups in the future. Conversely, community 23 has eight nodes, three of which have a default history. They could be merged with the neighboring communities. (3) Structural hole spanner nodes should be given special attention. Usually, defaults happen on the structural hole spanners, so the adjacent communities can be merged. Finally, we obtain ten communities, seven of which have relatively high default rates as \autoref{coitable} and as \autoref{merged}. The seven medium-sized groups of subgraphs can be efficiently processed for further tasks. It is noted that the merge and reassign operations are based on user expertise and the user may choose various criteria, the final treemap can demonstrate different combinations and default rates.

\begin{table}[tbp]\vspace{-5pt}
\centering
\tiny
\begin{tabular}{|l|c|c|c|c|c|c|c|c|c|c|}
\hline
Community ID & 13  & 12   & 3   &  6  & 7    & 8  &  9   \\ \hline
Firms    & 46   & 36   & 103  & 44   & 88   & 25   & 128  \\ \hline
Defaults    & 6    & 11   & 37   & 17   & 18   & 7    & 25   \\ \hline
Ratio for default firms & 13\% & 31\% & 36\% & 39\% & 20\% & 28\% & 20\% \\ \hline
Ratio for default amount & 31\% & 97\% & 85\% & 41\% & 40\% & 78\% & 51\% \\ \hline
Structural hole spanner   & 4    & 1    & 17   & 3    & 7    & 1    & 5    \\ \hline
Neighbour communities    & 2    & 1    & 6    & 1    & 2    & 1    & 2    \\ \hline
Total loan amount    & 623  & 826  & 1695 & 2080 & 2273 & 512  & 4045 \\ \hline
Total default amount    & 191  & 804  & 1441 & 863  & 918  & 398  & 2083 \\ \hline
\end{tabular}
\caption{Statistics for communities after interactive editing.}\vspace{-22pt}
\label{coitable}
\end{table}

\subsection{Risk Guarantee Patterns Discovery}\vspace{-2pt}

With the high default groups, we are able to focus and explore risk patterns more efficiently. This includes (1) automatic motif detection from high default groups. Specifically, we employ the gtrieScanner (http://www.dcc.fc.up.pt/gtries/) approach. (2) Matching the motifs with the entire network and calculating the ratio for default firms. (3) Ranking the motifs in descending default order, and they are high default patterns. (4) The user interactively edits the high default patterns by adding more nodes, and the system will automatically match the new subgraph with the entire network and produce the ratio for default firms.

\begin{figure*}[tb!]
  \centering
  \includegraphics[width=1\linewidth]{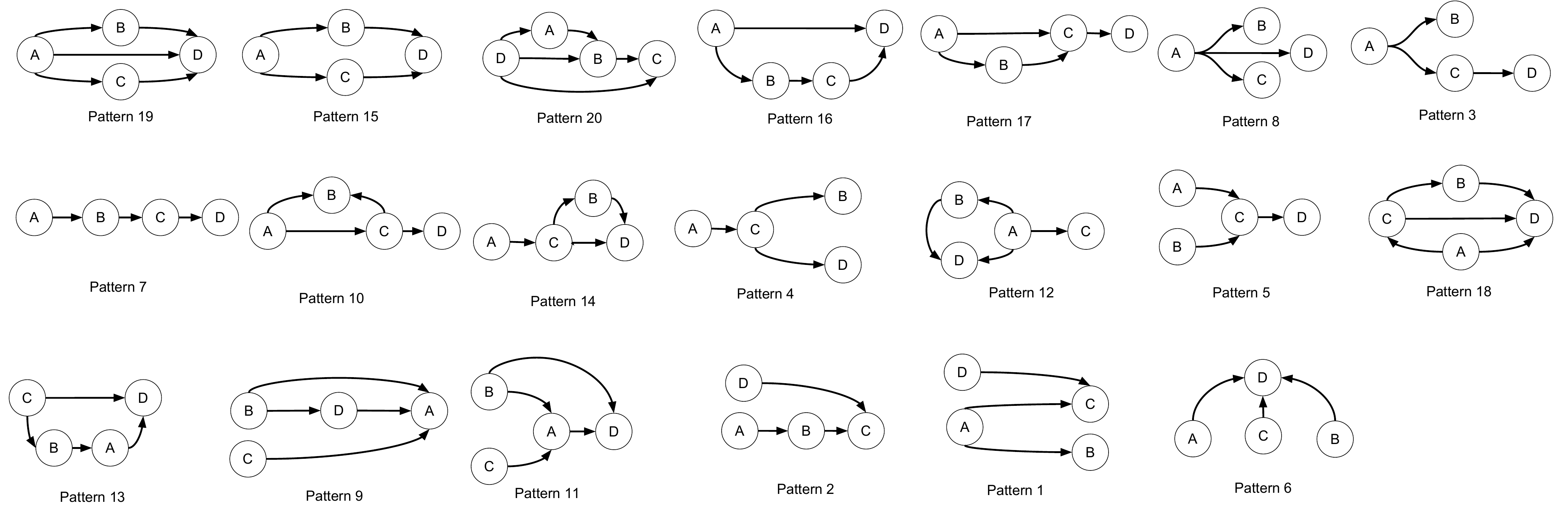}\vspace{-15pt}
  \caption{All the patterns (4-vertex-motif structures) detected from community 3. Among them, patterns 15, 16, and 17 show single-input, single-output, and feed-forward structures.}\label{reform}\vspace{-15pt}
\end{figure*}

Matching all those motifs on the whole network would be time-consuming. Theoretically, there are 199 and 9364 possible combinations for 4- and 5-vertex motifs for a directed network, respectively. We start from the 4-vertex-motifs and by interactively editing risk motifs, the user can explore more complex patterns efficiently. In the case study, we choose to analyze community 3, which consists of 103 enterprises; 36\% of them default the 85\% loans from the bank, as \autoref{coitable} shows. \autoref{reform} gives the twenty 4-vertex-motifs automatic algorithm detected from community 3, and \autoref{riskmotiftable} shows the statistical information.
\begin{table*}[tb!]\vspace{-10pt}
\tiny
\resizebox{0.96\textwidth}{!}{
\begin{tabular}{|l|c|c|c|c|c|c|c|c|c|c|c|c|c|c|c|c|c|c|c|c|}
\hline
Motif ID      & 19    & 15    & 20    & 16   & 17   & 8    & 3    & 7    & 10   & 14   & 4    & 12   & 5    & 18   & 13   & 11   & 2    & 1     & 6     & 9    \\ \hline
Motifs         & 1     & 4     & 1     & 4    & 4    & 74   & 169  & 92   & 23   & 6    & 164  & 17   & 151  & 1    & 13   & 22   & 312  & 437   & 95    & 24   \\ \hline
Firms         & 4     & 10    & 4     & 28   & 18   & 165  & 238  & 179  & 125  & 24   & 202  & 101  & 304  & 25   & 106  & 138  & 410  & 522   & 478   & 176  \\ \hline
default firms      & 4     & 9     & 3     & 18   & 11   & 79   & 110  & 69   & 48   & 9    & 70   & 35   & 89   & 7    & 28   & 32   & 95   & 111   & 79    & 26   \\ \hline
Ratio for default firm& 100 & 90  & 75  & 64 & 61 & 48 & 46 & 39 & 38 & 38 & 35 & 35 & 29 & 28 & 26 & 23 & 23 & 21  & 17  & 15 \\ \hline
Ratio for default amount      & 100 & 100 & 100 & 55 & 75 & 56 & 53 & 71 & 45 & 47 & 59 & 37 & 58 & 24 & 31 & 49 & 64 & 49  & 46  & 44 \\ \hline
Total loan amount          & 36    & 78    & 64    & 955  & 218  & 3259 & 5442 & 3583 & 3602 & 263  & 4872 & 3157 & 6975 & 1364 & 3134 & 4930 & 8919 & 11963 & 10546 & 3433 \\ \hline
Total default amount         & 36    & 78    & 64    & 522  & 163  & 1829 & 2897 & 2547 & 1607 & 123  & 2871 & 1166 & 4072 & 331  & 970  & 2405 & 5686 & 5822  & 4836  & 1507 \\ \hline
\end{tabular}}
\caption{Statistical information for the high default motifs.}\vspace{-15pt}
\label{riskmotiftable}
\end{table*}

\begin{figure}[tbh] \vspace{-10pt}
  \centering
  \includegraphics[width=1\linewidth]{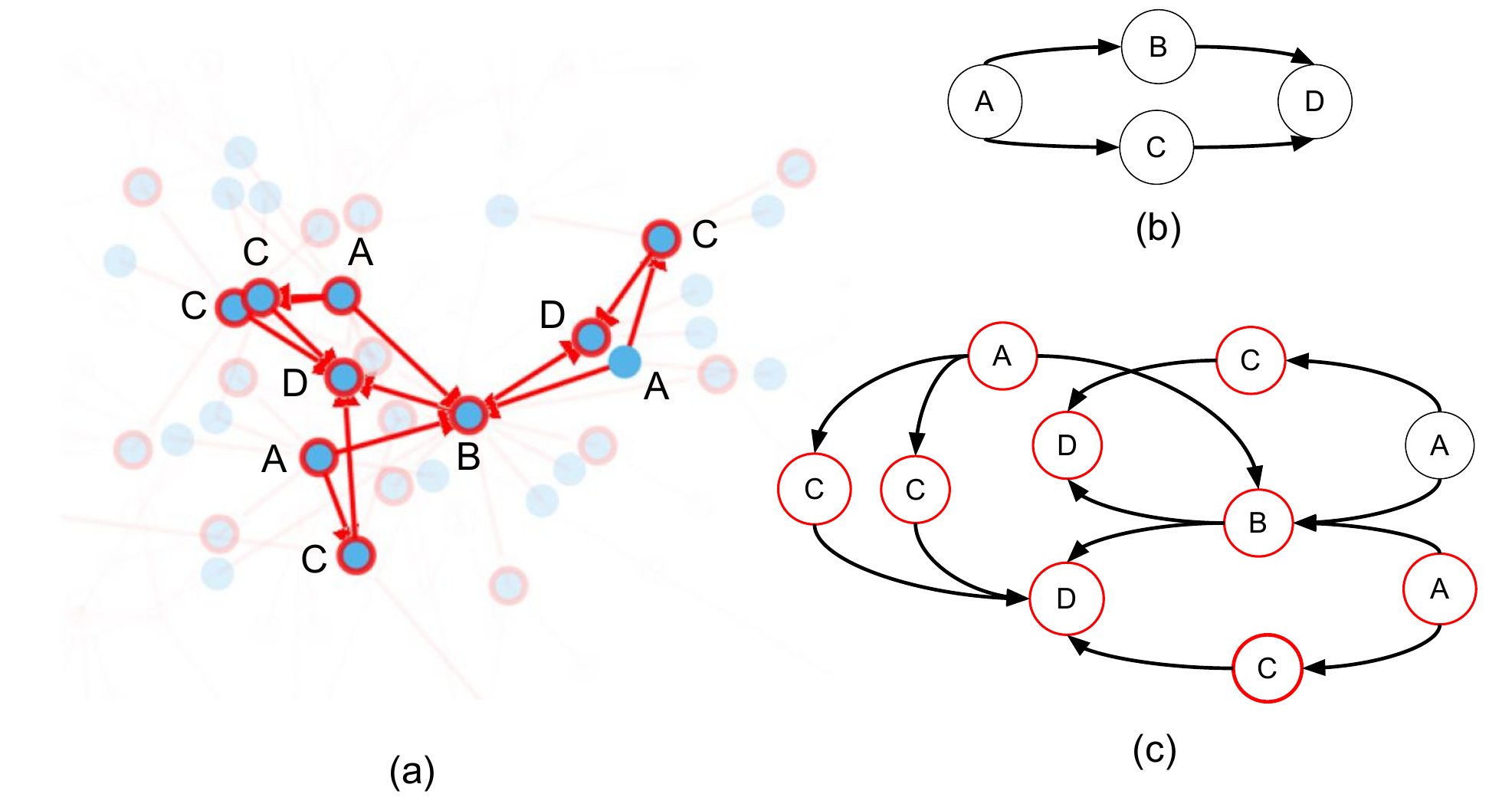}\vspace{-10pt}
  \caption{(a) Pattern 15 highlighted on the loan guarantee network. (b) Pattern 15 model. (c) Alternative way to understand pattern 15.}\label{M151617}\vspace{-13pt}
\end{figure}

Although there are nearly 200 kinds of 4-vertex node motif shapes, only 20 exist in the high default group. We thus perform analysis only on the 20 motifs rather than on every shape. Most of them have rather complex structures; however, some of them are known to banking experts -- for example, motif 6 is a joint liability loan. Some others can be understood by a combination of smaller guarantee patterns. For example, motif 5 is a combination of a joint liability guarantee loan with a single guarantee. Three of the motifs, 15, 16, and 17, attracted our attention for a number of reasons: (1) high default rates for the patterns (ranging from 61\% to 90\% in the ratio for default firm and 55\% to 100\% in the ratio for default amount); (2) a relatively small number of instances (4 or 5) are detected from the whole network; (3) the top five risk motifs show \emph{single input, single output, feed forward structures}. \autoref{M151617} gives all the instances of pattern 15 that are detected from the entire network. Some of the nodes coincide together. These three patterns are interesting; for example, where pattern 15 occurs five times in a group, the bank lost \emph{all} the money lent to the enterprises with such guarantee structures (see Table~\ref{riskmotiftable}). There is a high possibility that a fraud loan guarantee may happen several times, and the local bank failed to recognize the fraud pattern. A similar analysis implies that patterns 16 and 17 may also be risk patterns.

\section{User study}\vspace{-2pt}
We conduct interviews with two banking loan experts. Expert-A comes from the financial regulator. He has more than five years of experience in guarantee network research and has published several important investigation reports and books on the topic. He is also the expert who together with us to consolidate the four research tasks. Expert-B comes from our cooperated bank. He has ten years of loan approval experience and is able to access the complete data set. Both interviewees are attracted by and immediately understand the force-directed graph based view, however, they have difficult to further explore more functions. So, we give them both a 15 minutes training, introducing the main tasks, motivation and the operations of the interface. The interviewees could ask questions and operate the interface to warm up. Then, the interviewees are required to run the tool in 30 minutes and write down their feedback.\vspace{-2pt}

Expert-A is familiar with all the four tasks. In the Task 1, he agrees the indicators are useful but suggest the interface should be reorganized with buttons, as the current drop-down menu is a bit difficult to choose. In the Task 2, he is rather interested in the community editing. He said that when he and his colleagues try to resolve the financial risks in guarantee networks, a major operation is to split the loan guarantee network into smaller ones in order to avoid the default diffusion. The editing function of the tool provides users with a powerful weapon to achieve their target. He agrees on illegally conveyed benefits might exist under the suggested risk patterns. In the Task 3, he likes the animation but also suggested the dynamic information needs further investigation. In the Task 4, he suggested that the diffusion path and the corresponding Sankey diagram are useful, but better diffusion model should be developed. Finally, he suggested, the four sub-tools should be reorganized and integrated into one view so that that they can maximize the potential for the ultimate risk isolation operations. \vspace{-5pt}

Expert-B expressed that with the tool he is able to grasped the intricate connections between enterprises clearly when assessing a loan. In the Task 1, He likes the force-directed graph based monitoring view but expressed concern about visual clutter issue. He mentioned that in practice, the guaranteed network size could be as large as thousands nodes (although it is very rare) and in such case, it would be difficult to visualize them in the naive force-directed graph. In the Task 2, he thinks the treemap gives an intuitive understanding of the guarantee groups; he likes the financial radar view which we did not expect. He is very interested in the discovered risk patterns. He noted the ID of the nodes and investigated in-depth what had happened. Two weeks later, he sends us feedback that in pattern 15 (see \autoref{M151617}), all default enteritises were sued in court one after another a year ago. With the given names, we confirm that the enterprises are mostly in printing or related industry. However, because the businesses are very small (three to five employees on average), and the information is not transparent, we are not able to dig more. In the Task 3, he expressed that the animation is rather intuitive. It helps to understand how the network was generated but lacks a strong connection with other tools. In the Task 4, he understands the diffusion path and meaning of the  Sankey diagram. Because of the tool are currently could only analysis preloaded data, he suggests we further develop the interface and tests the tool on more data set.

\textbf{Discussion}. The above case studies and domain expert interviews confirm the effectiveness of the system in networked-guarantee loan risk management. We also notice there are some limitations. (1) Visual clutters. The case studies are performed on an independent subgraph with more than 600 nodes, the experts are able to zoom in to see the details on ordinary laptops without difficulty. In practice, extreme complex independent subgraphs are very rare and obey the power law~\cite{newman2003structure}. Our statics shows 85.1\% are graphs with fewer than 50 vertexes while about 6.6\% are graphs composed of more than 300 vertexes in a real dataset. So, the current system can be applied to the majority networked-guarantee loan risk management tasks. However, analyzing the large guarantee networks is also important, we believe classic graph simplification algorithms (for example, community-based clustering) may help to reduce the visual clutter and improve the visualize performance. (2) Visual interface optimization. The current system has separate sub-tool views for different tasks and the operation are relatively complex even for domain expert. Since we have conducted case study with domain experts, next, we will introduce it to the visual analytics experts and perform pair analytics. With the feedback, we will optimize the visual analytics work flow and the interface around risk isolation--the ultimate goal. (3) Default diffusion prediction.  All the vulnerable nodes are highlighted in the current system and it will be inevitably introduce misjudgment. Future work will include computational modeling of default diffusion.
\vspace{-5pt}
\section{Conclusion}\vspace{-2pt}
We present a visual analytics approach for networked-guarantee loan risk management. To our best knowledge, this is the first work using visual analytics approaches to address the guarantee loan default issue. It can help the government and banks to monitor default spread status and can provide insight for taking precautionary measures to prevent and dissolve systematic financial risk.
\vspace{-5pt}

\acknowledgments{
We would like to thank the anonymous reviewers for their useful feedbacks. This research was sponsored by the Open Project Program of the State Key Lab of CAD\&CG $($Grant No.A1824$)$, Zhejiang University and NVIDIA Corporation GPU Grant.}

%\bibliographystyle{abbrv}
%\bibliographystyle{abbrv-doi}
%\bibliographystyle{abbrv-doi-narrow}
%\bibliographystyle{abbrv-doi-hyperref}
%\bibliographystyle{abbrv-doi-hyperref-narrow}

%\bibliography{refe}

\begin{thebibliography}{10}

\bibitem{allen2010financial}
F.~Allen, A.~Babus, and E.~Carletti.
\newblock Financial connections and systemic risk.
\newblock Technical report, National Bureau of Economic Research, 2010.

\bibitem{archambault2011animation}
D.~Archambault, H.~Purchase, and B.~Pinaud.
\newblock Animation, small multiples, and the effect of mental map preservation
  in dynamic graphs.
\newblock {\em IEEE Transactions on Visualization and Computer Graphics},
  17(4):539--552, 2011.

\bibitem{archambault2016can}
D.~Archambault and H.~C. Purchase.
\newblock Can animation support the visualisation of dynamic graphs?
\newblock {\em Information Sciences}, 330:495--509, 2016.

\bibitem{baesens2003using}
B.~Baesens, R.~Setiono, C.~Mues, and J.~Vanthienen.
\newblock Using neural network rule extraction and decision tables for
  credit-risk evaluation.
\newblock {\em Management science}, 49(3):312--329, 2003.

\bibitem{baesens2003benchmarking}
B.~Baesens, T.~Van~Gestel, S.~Viaene, M.~Stepanova, J.~Suykens, and
  J.~Vanthienen.
\newblock Benchmarking state-of-the-art classification algorithms for credit
  scoring.
\newblock {\em Journal of the operational research society}, 54(6):627--635,
  2003.

\bibitem{battiston2012debtrank}
S.~Battiston, M.~Puliga, R.~Kaushik, P.~Tasca, and G.~Caldarelli.
\newblock Debtrank: Too central to fail? financial networks, the fed and
  systemic risk.
\newblock {\em Scientific reports}, 2:srep00541, 2012.

\bibitem{bisias2012survey}
D.~Bisias, M.~Flood, A.~W. Lo, and S.~Valavanis.
\newblock A survey of systemic risk analytics.
\newblock {\em Annu. Rev. Financ. Econ.}, 4(1):255--296, 2012.

\bibitem{bougheas2015complex}
S.~Bougheas and A.~Kirman.
\newblock Complex financial networks and systemic risk: A review.
\newblock In {\em Complexity and Geographical Economics}, pp. 115--139.
  Springer, 2015.

\bibitem{burt2004structural}
R.~S. Burt.
\newblock Structural holes and good ideas.
\newblock {\em American journal of sociology}, 110(2):349--399, 2004.

\bibitem{burt2007secondhand}
R.~S. Burt.
\newblock Secondhand brokerage: Evidence on the importance of local structure
  for managers, bankers, and analysts.
\newblock {\em Academy of Management Journal}, 50(1):119--148, 2007.

\bibitem{catanzaro2013network}
M.~Catanzaro and M.~Buchanan.
\newblock Network opportunity.
\newblock {\em Nature Physics}, 9:121--123, 2013.

\bibitem{chang2007wirevis}
R.~Chang, M.~Ghoniem, R.~Kosara, W.~Ribarsky, J.~Yang, E.~Suma, C.~Ziemkiewicz,
  D.~Kern, and A.~Sudjianto.
\newblock Wirevis: Visualization of categorical, time-varying data from
  financial transactions.
\newblock In {\em Visual Analytics Science and Technology, 2007. VAST 2007.
  IEEE Symposium on}, pp. 155--162. IEEE, 2007.

\bibitem{chen2016xgboost}
T.~Chen and C.~Guestrin.
\newblock Xgboost: A scalable tree boosting system.
\newblock In {\em Proceedings of the 22nd acm sigkdd international conference
  on knowledge discovery and data mining}, pp. 785--794. ACM, 2016.

\bibitem{dumas2014financevis}
M.~Dumas, M.~J. McGuffin, and V.~L. Lemieux.
\newblock Financevis. net: A visual survey of financial data visualizations.
\newblock In {\em Poster Abstracts of IEEE Conference on Visualization},
  vol.~2, 2014.

\bibitem{he2016joint}
L.~He, C.-T. Lu, J.~Ma, J.~Cao, L.~Shen, and P.~S. Yu.
\newblock Joint community and structural hole spanner detection via harmonic
  modularity.
\newblock In {\em Proceedings of the 22nd ACM SIGKDD International Conference
  on Knowledge Discovery and Data Mining}, pp. 875--884. ACM, 2016.

\bibitem{huang2009visualization}
M.~L. Huang, J.~Liang, and Q.~V. Nguyen.
\newblock A visualization approach for frauds detection in financial market.
\newblock In {\em Information Visualisation, 2009 13th International
  Conference}, pp. 197--202. IEEE, 2009.

\bibitem{jian2012determinants}
M.~Jian and M.~Xu.
\newblock Determinants of the guarantee circles: The case of chinese listed
  firms.
\newblock {\em Pacific-Basin Finance Journal}, 20(1):78--100, 2012.

\bibitem{khandani2010consumer}
A.~E. Khandani, A.~J. Kim, and A.~W. Lo.
\newblock Consumer credit-risk models via machine-learning algorithms.
\newblock {\em Journal of Banking \& Finance}, 34(11):2767--2787, 2010.

\bibitem{klukas2006coordinated}
C.~Klukas, F.~Schreiber, and H.~Schw{\"o}bbermeyer.
\newblock Coordinated perspectives and enhanced force-directed layout for the
  analysis of network motifs.
\newblock In {\em Proceedings of the 2006 Asia-Pacific Symposium on Information
  Visualisation-Volume 60}, pp. 39--48. Australian Computer Society, Inc.,
  2006.

\bibitem{lancichinetti2008benchmark}
A.~Lancichinetti, S.~Fortunato, and F.~Radicchi.
\newblock Benchmark graphs for testing community detection algorithms.
\newblock {\em Physical review E}, 78(4):046110, 2008.

\bibitem{lindskog2000modelling}
F.~Lindskog et~al.
\newblock {\em Modelling dependence with copulas and applications to risk
  management}.
\newblock PhD thesis, Master Thesis, ETH Z{\"u}rich, 2000.

\bibitem{maguire2013visual}
E.~Maguire, P.~Rocca-Serra, S.-A. Sansone, J.~Davies, and M.~Chen.
\newblock Visual compression of workflow visualizations with automated
  detection of macro motifs.
\newblock {\em IEEE transactions on visualization and computer graphics},
  19(12):2576--2585, 2013.

\bibitem{masters1982rasch}
G.~N. Masters.
\newblock A rasch model for partial credit scoring.
\newblock {\em Psychometrika}, 47(2):149--174, 1982.

\bibitem{mcmahon2014loan}
D.~Mcmahon.
\newblock Loan ‘guarantee chains’ in china prove flimsy.
\newblock {\em The Wall Street Journal}, 27, 2014.

\bibitem{meng2017netrating}
X.~Meng, Y.~Tong, X.~Liu, Y.~Chen, and S.~Tan.
\newblock Netrating: Credit risk evaluation for loan guarantee chain in china.
\newblock In {\em Pacific-Asia Workshop on Intelligence and Security
  Informatics}, pp. 99--108. Springer, 2017.

\bibitem{meng2015credit}
X.~L.~X. Meng.
\newblock Credit risk evaluation for loan guarantee chain in china.
\newblock 2015.

\bibitem{newman2003structure}
M.~E. Newman.
\newblock The structure and function of complex networks.
\newblock {\em SIAM review}, 45(2):167--256, 2003.

\bibitem{Zhibin2015Rapidly}
Z.~Niu, R.~R. Martin, F.~C. Langbein, and M.~A. Sabin.
\newblock Rapidly finding cad features using database optimization.
\newblock {\em Computer-Aided Design}, 69(C):35--50, 2015.

\bibitem{rosvall2008maps}
M.~Rosvall and C.~T. Bergstrom.
\newblock Maps of random walks on complex networks reveal community structure.
\newblock {\em Proceedings of the National Academy of Sciences},
  105(4):1118--1123, 2008.

\bibitem{rudolph2009finvis}
S.~Rudolph, A.~Savikhin, and D.~S. Ebert.
\newblock Finvis: Applied visual analytics for personal financial planning.
\newblock In {\em Visual Analytics Science and Technology, 2009. VAST 2009.
  IEEE Symposium on}, pp. 195--202. IEEE, 2009.

\bibitem{sarlin2011clustering}
P.~Sarlin.
\newblock Clustering the changing nature of currency crises in emerging
  markets: an exploration with self-organising maps.
\newblock {\em International Journal of Computational Economics and
  Econometrics}, 2(1):24--46, 2011.

\bibitem{sarlin2011sovereign}
P.~Sarlin.
\newblock Sovereign debt monitor: A visual self-organizing maps approach.
\newblock In {\em Computational Intelligence for Financial Engineering and
  Economics (CIFEr), 2011 IEEE Symposium on}, pp. 1--8. IEEE, 2011.

\bibitem{sarlin2013chance}
P.~Sarlin.
\newblock Chance discovery with self-organizing maps: Discovering imbalances in
  financial networks.
\newblock {\em Advances in Chance Discovery}, pp. 49--61, 2013.

\bibitem{von2015visual}
T.~von Landesberger, S.~Diel, S.~Bremm, and D.~W. Fellner.
\newblock Visual analysis of contagion in networks.
\newblock {\em Information Visualization}, 14(2):93--110, 2015.

\bibitem{von2009system}
T.~von Landesberger, M.~G{\"o}rner, R.~Rehner, and T.~Schreck.
\newblock A system for interactive visual analysis of large graphs using motifs
  in graph editing and aggregation.
\newblock In {\em VMV}, vol.~9, pp. 331--340, 2009.

\bibitem{Yan2015Multi}
J.~Yan, M.~Cho, H.~Zha, X.~Yang, and S.~M. Chu.
\newblock Multi-graph matching via affinity optimization with graduated
  consistency regularization.
\newblock {\em IEEE Transactions on Pattern Analysis \& Machine Intelligence},
  38(6):1228--1242, 2015.

\end{thebibliography}
\end{document}